\documentclass[12pt,preprint]{aastex}

\newcommand{\etal }{{et al.} }
\newcommand{\msun}{\thinspace M_\odot}

\newcommand{\vect}[1]{\mbox{\boldmath$#1$}}
\def\lesssim{\mathrel{\hbox{\rlap{\hbox{\lower4pt\hbox{$\sim$}}}\hbox{$<$}}}}
\def\gtrsim{\mathrel{\hbox{\rlap{\hbox{\lower4pt\hbox{$\sim$}}}\hbox{$>$}}}}
\newcommand{\cm}{\,{\rm cm}^{-3} } 
\newcommand{\jj}{{\rm g}\,{\rm cm}^{2}\, \rm{s}^{-1} } 
 
\newcommand{\nc}{n_{\rm c} }

\newcommand{\tc}{t_{\rm c}}

\newcommand{\dfrac}[2]{{\displaystyle \frac{#1}{#2}} }

\shorttitle{Circumstellar Disk Formation}
\shortauthors{Machida \etal 2010}

\begin{document}
\title{Effect of Magnetic Braking on the Circumstellar Disk Formation in a Strongly Magnetized Cloud}
\author{Masahiro N. Machida\altaffilmark{1}, Shu-ichiro Inutsuka\altaffilmark{2}, and Tomoaki Matsumoto\altaffilmark{3}} 
\altaffiltext{1}{National Astronomical Observatory of Japan, Mitaka, Tokyo 181-8588, Japan; masahiro.machida@nao.ac.jp}
\altaffiltext{2}{Department of Physics Nagoya University Furo-cho, Chikusa-ku Nagoya, Aichi 464-8602; inutsuka@nagoya-u.jp}
\altaffiltext{3}{Faculty of Humanity and Environment, Hosei University, Fujimi, Chiyoda-ku, Tokyo 102-8160, Japan; matsu@hosei.ac.jp}

\begin{abstract}
Using resistive magnetohydrodynamics simulation, we investigate circumstellar disk formation in a strongly magnetized cloud.
As the initial state, an isolated cloud core embedded in a low-density interstellar medium with a uniform magnetic field is adopted.
The cloud evolution is calculated until almost all gas inside the initial cloud falls onto either the circumstellar disk or a protostar, and a part of the gas  is ejected into the interstellar medium by the protostellar outflow driven by the circumstellar disk.
In the early main accretion phase, the disk size is limited to $\sim$10\,AU because the angular momentum of the circumstellar disk is effectively transferred by both magnetic braking and the protostellar outflow.
In the later main accretion phase, however, the circumstellar disk grows rapidly and exceeds  $\gtrsim100$\,AU by the end of the main accretion phase.
This rapid growth of the circumstellar disk is caused by the depletion of the infalling envelope, while magnetic braking is effective when the infalling envelope is more massive than the circumstellar disk.
The infalling envelope cannot brake the circumstellar disk when the latter is more massive than the former.
In addition, the protostellar outflow weakens and disappears in the later main accretion phase, because the outflow is powered by gas accretion onto the circumstellar disk.
Although the circumstellar disk formed in a magnetized cloud is considerably smaller than that in an unmagnetized cloud, a circumstellar disk exceeding $100$\,AU can form even in a strongly magnetized cloud.
\end{abstract}

\section{Introduction}
\label{sec:intro}
The magnetic field is an essential ingredient in present-day star formation.
Observations have shown that molecular cloud cores have magnetic energy comparable to gravitational energy \citep{crutcher99}.
The cloud cores also have angular momentum \citep{arquilla86,goodman93,caselli02}.
Stars are born in such cloud cores. 
As the cloud collapses, the central region of the collapsing cloud rotates rapidly, and this rapid rotation suppresses further collapse and subsequent protostar formation.
In the collapsing cloud, however, the magnetic field can effectively transfer the excess angular momentum from the center of the cloud by magnetic braking (and protostellar outflow), and promote further collapse and protostar formation.
Moreover, after the protostar formation, the magnetic field greatly contributes to the evolution of a protostar and a circumstellar disk.
Because the angular momentum in the circumstellar disk is transferred by magnetic effects, the gas in the circumstellar disk can fall onto the protostar, increasing the protostellar mass in the main accretion phase.
Thus, it may be expected that only a small-size (or no) disk appears around the protostar if the angular momentum of the disk is effectively transferred by magnetic effects.
In various star-forming regions, however, many circumstellar disks with a size of several hundred AU can be observed (e.g., \citealt{calvet00,natta00}), indicating that some of the angular momentum survives the magnetic braking catastrophe and contributes to the growth of the circumstellar disk.

The formation and evolution of the circumstellar disk are important not only for star formation but also for planet formation.
The size and mass of the circumstellar disk determine the mode of formation of a planet (i.e., core accretion or gravitational instability mode; see \citealt{durisen07}).
%% ( \citealt{durisen07}).
The angular velocity of the molecular cloud core that is the host cloud for star formation creates a circumstellar disk in the star formation process.
Thus, to understand the formation of the circumstellar disk, we need to investigate the evolution of the molecular cloud core from the prestellar stage.
Numerical simulations can reproduce the disk formation in the molecular cloud core.
However, because there is a large difference in both spatial and time scales between the molecular cloud core and the circumstellar disk, we need special numerical techniques such as adaptive mesh refinement or nested grid and smoothed particle hydrodynamics.
Using these techniques, several studies have investigated the disk formation in the molecular cloud core from the prestellar stage.
\citet{bate10} and \citet{machida10b} investigated the formation of the protostar and circumstellar disk in an unmagnetized cloud; they showed that the first core that forms before protostar formation in the gas-collapsing phase directly evolves into a circumstellar disk after the protostar formation.
\citet{walch09a} and \citet{machida10a} showed that the circumstellar disk, which originates from the first core, is more massive than or comparable to the protostar in the main accretion phase and has a size of several hundred AU when the host cloud is unmagnetized.

Using numerical simulations, the disk formation in magnetized clouds has also been investigated.
These studies reported that the calculated size of the circumstellar disk appearing in a magnetized cloud is much smaller than the observed sizes.
%% which is recognized as a serious problem in star formation process.
%%These studies showed that the size of the circumstellar disk appeared in the magnetized cloud is much smaller than those of observations, which is recognized as serious problem in the star formation.
Using an ideal MHD calculation, \citet{mellon08} showed that the circumstellar disk formed in a magnetized cloud has a size of $\sim10$\,AU at maximum.
\citet{hennebelle09} also showed that large-size disk formation is suppressed in a magnetized cloud.
This is because the angular momentum of the circumstellar disk (or infalling gas onto the circumstellar disk) is excessively transferred outward by magnetic braking, and the gas in the disk effectively falls onto the protostar.
In reality, however, the magnetic field dissipates in a high-density gas region by Ohmic dissipation and ambipolar diffusion, because the degree of ionization  is considerably low (e.g., \citealt{nakano02}).
Such a high-density gas region, where the neutral gas is decoupled from the magnetic field, corresponds well to the disk-forming region in the early main accretion phase ($\sim0.1-10$\,AU, \citealt{machida07,machida08a,inutsuka10}).
%%($\sim0.1-10$\,AU, \citealt{machida07,machida08a}).  
Thus, if the magnetic field can sufficiently dissipate around the protostar, magnetic braking becomes ineffective and a large-size circumstellar disk comparable to observations may appear even in strongly magnetized clouds.
%%In contrast, with MHD calculation with different prescription for non-ideal effect
In contrast, with non ideal MHD calculations, \citet{mellon09} and \citet{duffin09} reported that the disk size is considerably smaller than observations  even when the magnetic field dissipates in the high-density gas region, although the  disk in non ideal MHD calculations is somewhat larger than that in ideal MHD calculations.

These studies, however, overlooked the effect of the finite mass of the host cloud for star and circumstellar disk formation. 
In the early main accretion phase, a massive gas envelope (or remnant of the molecular cloud) remains around the circumstellar disk.
The massive envelope rotates much slower than the circumstellar disk, and thus it can brake the circumstellar disk through the magnetic field lines that connect the less massive rapidly rotating disk to the more massive slowly rotating envelope.
In other words, the angular momentum of the circumstellar disk is transferred into the infalling envelope by torsional Alfv$\acute{\rm e}$n waves.
This is the magnetic braking mechanism for the circumstellar disk formation.
Thus, even with very strong magnetic fields, the circumstellar disk rotation velocity does not decrease when no massive envelope exists around the disk.
The mass of the infalling envelope decreases with time in the main accretion phase when an isolated molecular cloud core is assumed. 
The infalling envelope disappears by the end of the main accretion phase: a part of the envelope is blown away from the cloud by the protostellar outflow \citep{matzner00,machida09a}, and the remainder falls onto either the circumstellar disk or the protostar.
Thus, we can expect that magnetic braking becomes gradually ineffective as the mass of the infalling envelope decreases, and it rarely works when the mass of the circumstellar disk is much larger than that of the infalling envelope.
Note that, in reality,  magnetic braking can work even after the infalling envelope disappears, because the low-density interstellar medium can receive angular momentum from the circumstellar disk.
Also note that because the efficiency of magnetic braking decreases as the ambient density decreases, the angular momentum of the circumstellar disk is not effectively transferred into the low-density interstellar medium (see, \S\ref{sec:diss}).

In addition, in an isolated cloud core, although magnetic braking can redistribute the angular momentum inside the cloud, it cannot effectively transfer the angular momentum outside the cloud (or into the interstellar medium).
The infalling envelope receives the angular momentum from the circumstellar disk by magnetic braking, while it finally falls onto the circumstellar disk with the angular momentum by the end of the main accretion phase.
However, when the infalling envelope is depleted and  magnetic braking becomes ineffective, no sufficiently large disk appears if insufficient mass to form the disk remains in the infalling envelope.
%%the infalling envelope remains no sufficient mass to form the disk.

In summary, to investigate the real size and mass of the circumstellar disk, we have to investigate the disk evolution in the collapsing cloud core, at least till the end of the main accretion phase (or until the infalling gas disappears).
In previous studies, the calculation was stopped much before the infalling envelope disappears.
Thus, it is natural that magnetic braking is effective and no large-size disk appears during this phase (i.e., early main accretion phase).
In this study, however, we calculated the evolution of the circumstellar disk until almost all gas of the infalling envelope disappears and found that a large-size disk comparable to observations appears when the infalling envelope becomes less massive than the circumstellar disk.
The structure of this paper is as follows. 
The framework of our models and the numerical method are given in \S 2. 
The numerical results are presented in \S 3. 
We discuss the magnetic braking timescale and the effect of the infalling envelope on the circumstellar disk formation in \S 4.

\section{Model Settings}
\subsection{Basic Equations and Initial State}
\label{sec:initial}
To study the formation and evolution of a circumstellar disk in a magnetized molecular cloud core, we solve the three-dimensional resistive MHD equations including self-gravity:
\begin{eqnarray} 
& \dfrac{\partial \rho}{\partial t}  + \nabla \cdot (\rho \vect{v}) = 0, & \\
& \rho \dfrac{\partial \vect{v}}{\partial t} 
    + \rho(\vect{v} \cdot \nabla)\vect{v} =
    - \nabla P - \dfrac{1}{4 \pi} \vect{B} \times (\nabla \times \vect{B})
    - \rho \nabla \phi, & \\ 
& \dfrac{\partial \vect{B}}{\partial t} = 
   \nabla \times (\vect{v} \times \vect{B}) + \eta \nabla^2 \vect{B}, & 
\label{eq:reg}\\
& \nabla^2 \phi = 4 \pi G \rho, &
\end{eqnarray}
where $\rho$, $\vect{v}$, $P$, $\vect{B} $, $\eta$, and $\phi$ denote density, velocity, pressure, magnetic flux density, resistivity, and gravitational potential, respectively.
To mimic the thermal evolution calculated by \citet{masunaga00}, we adopt the piece wise polytropic equation of state (see also \citealt{vorobyov06,machida07,mellon08, mellon09,inutsuka10}) as 
\begin{equation} 
P =  c_{s,0}^2\, \rho \left[ 1+ \left(\dfrac{\rho}{\rho_c}\right)^{2/3} \right],
\label{eq:eos}
\end{equation}
 where $c_{s,0} = 190$\,m\,s$^{-1}$ and 
$ \rho_c = 3.84 \times 10^{-14} \, \rm{g} \, \cm$ ($n_c = 10^{10} \cm$). 
With equation~(\ref{eq:eos}), the gas behaves isothermally for $n\lesssim10^{10}\cm$ and adiabatically for $n\gtrsim10^{10}\cm$.
For a realistic evolution of the magnetic field in the circumstellar disk, we adopt the resistivity $\eta$ as the fiducial value in \citet{machida07}, in which the Ohmic dissipation becomes effective for $10^{11}\cm \lesssim n \lesssim 10^{15}\cm$ (for details, see Eqs.~[9] and [10], and Fig.~1 of \citealt{machida07}).
%%(for details, see Eqs.~[9] and [10], and Fig.~1 of \citealt{machida07}).

\begin{figure}
\begin{center}
\includegraphics[width=150mm]{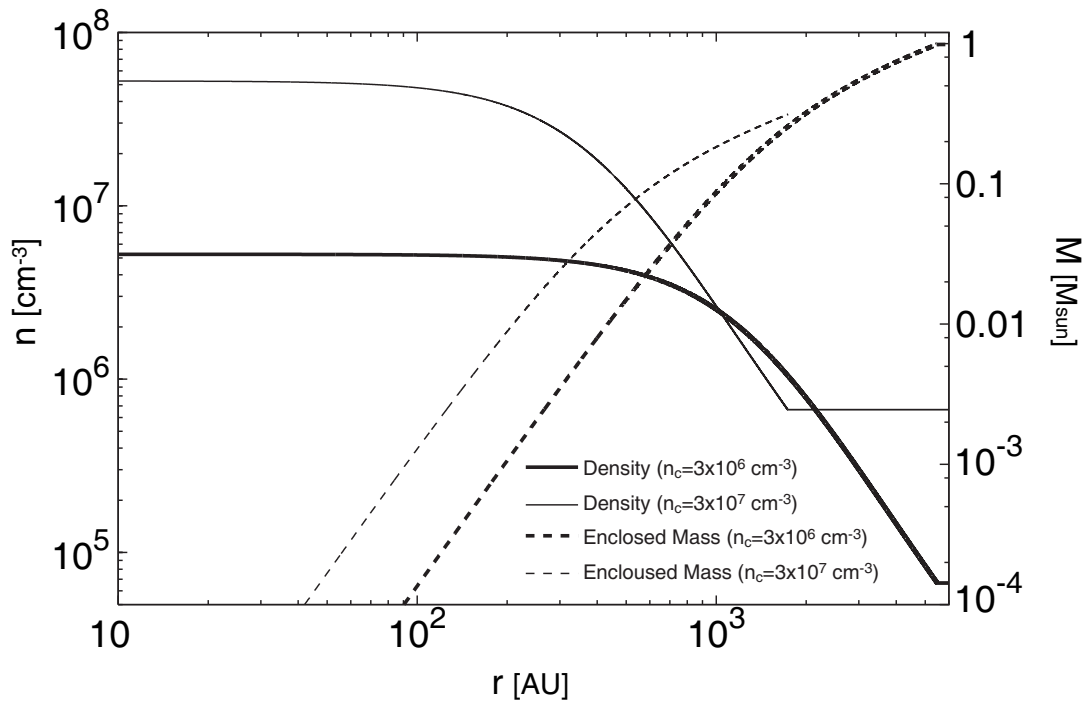}
\end{center}
\caption{Density distribution ({\it left axis}) and enclosed mass ({\it right axis}) are plotted against the initial cloud radius for models with a central density $n_{\rm c,0}=3\times 10^6\cm$ (thick line) and $3\times10^7\cm$ (thin line).
Only the enclosed mass in the range of $r<2\,r_{\rm BE}$ is plotted.
}
\label{fig:1}
\end{figure}

As the initial state, we take a spherical cloud with a Bonnor--Ebert (BE) density profile, which is widely recognized as the density profile of a prestellar cloud \citep{alves01,kandori05}.
The initial cloud  extends up to $ r= 2\, r_{\rm BE}$, where $r_{\rm BE}$ is the critical BE radius.
Outside the initial cloud ($r>2\, r_{\rm BE}$), a uniform density is adopted to mimic the static interstellar medium.
We add nonaxisymmetric density perturbation of the m = 2 mode to the initial core (for details, see \citealt{machida10a}).
%%(for details, see \citealt{machida10a}).
This perturbation ensures that the center of gravity is always located at the origin.
We ignore gravity (self-gravity and gravity of the protostar) in the region of $r>2\, r_{\rm BE}$.
In addition, we prohibit the gas inflow at $r=2\,r_{\rm BE}$ in order to suppress mass input from the interstellar medium into the gravitationally collapsing cloud core.
Thus, only the gas inside $r < 2\, r_{\rm BE}$ collapses to form the circumstellar disk and protostar.
Note that although the protostellar outflow driven by the circumstellar disk propagates into the region of $r > 2\,r_{\rm BE}$ and disturbs the interstellar medium over time, we can safely calculate the mass accretion process onto the circumstellar disk because the outflow propagating into the interstellar medium does not affect the inner cloud region  ($r < 2\, r_{\rm BE}$).
For the BE density profile, we adopt the central density as $n_{\rm c,0} =  3 \times 10^{4} \cm$, $ 3 \times 10^{6} \cm$ (fiducial model), and  $ 3 \times 10^{7} \cm$  with an isothermal temperature of $T=10$\,K.
To promote contraction, we increase the density by a factor of $f$=1.68, where $f$ is the density enhancement factor that represents the stability of the initial cloud.
With $f=1.68$, the initial cloud has $E_{\rm th}/\vert E_{\rm grav}\vert=0.5$ for $r<r_{\rm BE}$ and 0.42 for $r<2\,r_{\rm BE}$, where $E_{\rm th}$ and $E_{\rm grav}$ are the thermal and gravitational energy, respectively.
Although the central density of the initial cloud has values of $n_{\rm c,0}=3\times10^4 f \cm$, $3\times10^6 f \cm $ and $3\times10^7 f \cm$, we describe the initial cloud density without the density enhancement factor $f$ as $n_{\rm c,0}=3\times10^4\cm$ in the following.
%%Although the initial cloud has $n_{\rm c,0}=3\times10^4 f \cm$, $3\times10^6 f \cm $ and $3\times10^7 f \cm$ of the central density, respectively,  we describe the initial cloud density without the density enhancement factor $f$ as $n_{\rm c,0}=3\times10^4\cm$ in the following.
Note that because the critical BE radius (or initial cloud size) is determined by the central density without $f$, the initial density without $f$ is a more suitable parameter for describing the initial cloud.
The density and enclosed mass against the initial cloud radius are plotted in Figure~\ref{fig:1} for the model with $n_{\rm c,0} =  3 \times 10^{6} \cm$ and $ 3 \times 10^{7} \cm$.
The density contrast between the center of the cloud ($r=0$) and cloud boundary ($r=2\,r_{\rm BE}$) is $\rho (r=0)/\rho (r=2\,r_{\rm BE})=82$.
The cloud with a central density of $n_{\rm c,0} = 3\times10^{6} \cm$ (fiducial model) has a radius of $r_{\rm cl}= 5400$\,AU and mass of $M_{\rm cl}=1\msun$.
On the other hand, the cloud with $n_{\rm c,0} = 3\times10^{4} \cm$ and $3\times10^{7} \cm$ have a radius of $r_{\rm cl}= 2\,r_{\rm BE} =5.4\times10^4$\,AU and $1740$\,AU, and a mass of $M_{\rm cl}=10\msun$ and $0.3\msun$, respectively.
As seen in Figure~\ref{fig:1}, because we adopted the BE density profile, the density has a nearly flat distribution around the center of the cloud, while it drops in proportion to $r^{-2}$ in the outer envelope.

The cloud rotation generates a rotating disk around the protostar as the cloud collapses.
However, it is considered that the magnetic field suppresses the formation of the rotationally supported  disk in the collapsing cloud, because the magnetic field effectively transfers the angular momentum outward from the center of the  cloud.
In this study, to investigate the disk formation in an initially rotating magnetized cloud, we parameterized the magnetic field strength and rotation rate of the initial cloud.
We adopt a uniform magnetic field along the $z$-axis in the whole computational domain and rigid rotation along the $z$-axis in the region of $r<2\,r_{\rm BE}$.
Thus, the rotation axis is parallel to the magnetic field lines at the initial state.
To specify the initial cloud, we use two parameters describing the magnetic field strength ($\alpha$) and rotation rate ($\omega$).
They are normalized by the central quantities and described as 
\begin{equation}
\alpha = \dfrac{B_{\rm z,0}^2}{4\pi\,\rho_{\rm c,0} c_{\rm s,0}^2},
\end{equation}
where $B_{\rm z,0}$ is the magnetic flux density, and 
\begin{equation}
\omega = \dfrac{\Omega_{\rm c,0}}{\sqrt{4\pi\,{\rm G}\, \rho_{\rm c,0}}},
\end{equation}
where $\Omega_{\rm c,0}$ is the angular velocity at the center of the cloud.
Model names and their parameters are listed in Table~\ref{table:1}.
We adopt $\alpha =$ 0.1, 0.3, and 1, and $\omega=0.05$, 0.1 and 0.15, as described in Table~\ref{table:1}.

\begin{figure}
\begin{center}
\includegraphics[width=140mm]{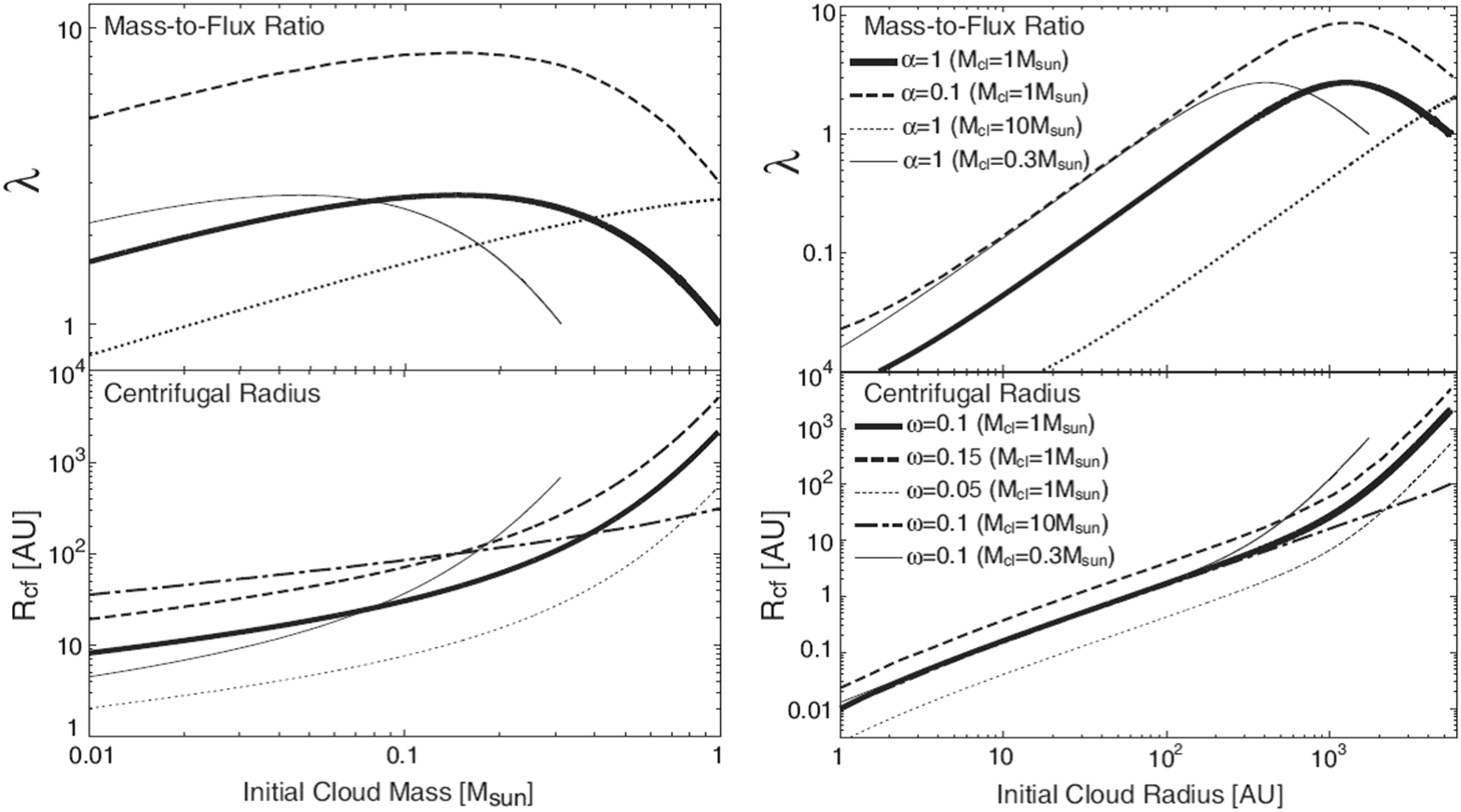}
\end{center}
\caption{
Initial mass-to-flux ratios normalized by the critical value ($\lambda$) for models with different initial magnetic field strengths $\alpha$ or different cloud masses are plotted against the initial cloud mass (left panel) and cloud radius (right panel) in the upper panels.
The centrifugal radii ($R_{\rm cf}$) for different angular velocities $\omega$ or different cloud masses are plotted in the lower panels.
}
\label{fig:2}
\end{figure}

In this study, we investigate disk formation until almost all the mass of the initial cloud falls onto either the circumstellar disk or the protostar.
Note that a part of the cloud mass is ejected from the collapsing cloud by protostellar outflow.
Thus, quantities describing the whole cloud  may be useful to specify our models.
The parameter $\alpha$ (or the magnetic flux density at the center of the cloud) can be connected to the mass-to-flux ratio of the whole cloud.
The mass-to-flux ratio is described as
\begin{equation}
\dfrac{M(r)}{\Phi(r)} = \frac{M(r)}{\pi r^2 B_0},
\label{eq:mag1}
\end{equation}
where $M(r)$ is the mass contained within the radius $r$ and  $\Phi(r)$ is the magnetic flux threading the cloud.
There exists a critical value of $M/\Phi$ below which a cloud is supported against  gravity by the magnetic field.
For a cloud with uniform density,  \citet{mouschovias76} derived a critical mass-to-flux ratio
\begin{equation}
\left(\dfrac{M}{\Phi}\right)_{\rm cri} = \dfrac{\zeta}{3\pi}\left(\dfrac{5}{G}\right)^{1/2},
\label{eq:mag2}
\end{equation}
where the constant $\zeta=0.48$ \citep{tomisaka88a,tomisaka88b}.
The mass-to-flux ratio normalized by the critical value $\lambda$ is described as
\begin{equation}
\lambda (r) \equiv \left(\dfrac{M(r)}{\Phi(r)}\right) \left(\dfrac{M}{\Phi}\right)_{\rm cri}^{-1}.
\label{eq:crit}
\end{equation}
The mass-to-flux ratios normalized by the critical value for different models (or different $\alpha$) are plotted against the enclosed mass ({\it left}) and the radius ({\it right}) of the initial cloud in the upper panels of Figure~\ref{fig:2}.
Reflecting the BE density profile, the mass-to-flux ratio has a radial distribution inside the initial cloud.
For example, with $\alpha=1$, the mass-to-flux ratio has a peak at $\sim0.1\msun$ ($\sim0.05\msun$) or $\sim10^3$\,AU ($\sim400$\,AU) for the model with $n_{\rm c,0}=3\times 10^6  \cm$ ($n_{\rm c,0}=3\times 10^7  \cm$).
In addition, with $\alpha=1$ and $n_{\rm c,0}=3\times 10^6\cm$, the initial cloud has a mass-to-flux ratio of  $\lambda(5400\,{\rm AU}) \simeq1$ as the whole.
Hereafter, for simplicity, we describe the mass-to-flux ratio of the whole cloud ($r<2\,r_{\rm BE}$) as $\lambda$ ($\equiv \lambda [2\,r_{\rm BE})])$, and call it the mass-to-flux ratio.
As listed in Table~\ref{table:1}, we adopt $\lambda=1$, 2, 3 (magnetized model) and $\infty$ (unmagnetized model).
The observations indicate that molecular cloud cores have the mass-to-flux ratio in the range of $0.8\lesssim \lambda \lesssim7.2$ with a median value of $\lambda \approx 2$ \citep{crutcher99}.
Thus, the mass-to-flux ratios adopted in this paper are within a reasonable range.

As the parameter for the rotation of the whole cloud, we define the initial ratio of the rotational to the gravitational energy as 
\begin{equation}
\beta_0 \equiv \dfrac{ E_{\rm r} }{ \vert E_{\rm g} \vert },
\end{equation}
where $E_{\rm r}$ is the rotational energy of the initial cloud in the region of $r<2\,r_{\rm BE}$.
As listed in Table~\ref{table:1},  models have the parameter in the range of $0.007\le\beta_0\le0.04$.
Observations showed that molecular cloud cores have the rotational energy in the range of $10^{-4} \lesssim \beta_0 \lesssim 0.07$ with a median value of $\beta_0 \approx 0.02$ \citep{goodman93,caselli02}.
Thus, the rotational energies in our models are also within reasonable ranges, compared with the observations.
In addition, we also listed the centrifugal radius of each model in Table~\ref{table:1}.
The centrifugal radius is defined as
\begin{equation}
r_{\rm cf}(r) = \dfrac{j(r)^2}{G\,M(r)},
\label{eq:cent}
\end{equation}
where $j ({\rm r})$ is the specific angular momentum.
The centrifugal radii for models with different $\omega$ or different $n_{\rm c,0}$ are plotted against the enclosed mass ({\it left}) and the radius ({\it right}) of the initial cloud in the lower panels of Figure~\ref{fig:2}.
In the figure, the centrifugal radius gradually increases as the enclosed mass increases, because the gas far from the center of the cloud has a large specific angular momentum.
As listed in Table~\ref{table:1}, models have the centrifugal radius in the range of $540{\rm AU}\le r_{\rm cf} \le 2.15\times 10^4\,{\rm AU}$ as the whole cloud ($r<2\,r_{\rm BE}$).
Thus, when the angular momentum is well conserved in the collapsing cloud, a disk with a size of $\gtrsim 1000$\,AU can form around the protostar.

\subsection{Numerical Method and Boundary Condition}
For calculations on a large spatial scale, the nested grid method is adopted (for details, see \citealt{machida05a,machida05b}). 
Each level of a rectangular grid has the same number of cells ($ 64 \times 64 \times 32 $).
The calculation is first performed with five grid levels ($l=1-5$).
The box size of the coarsest grid $l=1$ is chosen to be $L_1=2^5\, r_{\rm BE}$, which corresponds to $\sim 8.7\times 10^5$\,AU ($n_{\rm c,0}=3\times10^4  \cm$),  $\sim 8.7\times 10^4$\,AU ($n_{\rm c,0}=3\times10^6  \cm$) and $\sim 2.7\times 10^4$\,AU ($n_{\rm c,0}=3\times10^7  \cm$).
%% for large cloud ($\sim 5.6\times 10^4$\,AU for small cloud).
A new finer grid is generated before the Jeans condition is violated \citep{truelove97}.
For model with $n_{\rm c,0}=3\times10^6 \cm$, the maximum level of grids is $l_{\rm max} = 12$, which has a box size of $L_{12}=43$\,AU and a cell width of $\Delta_{l_{\rm max}}=0.67$\,AU.
Models with different cloud sizes (or different initial central densities) have almost the same spatial resolution ($\Delta_{l_{\rm max}}\sim0.5$\,AU) of the maximum level grid, in which the maximum grid level is different in models with different initial central density.

As described in \S\ref{sec:initial}, we adopt gravity only in the region of $r<2\,r_{\rm BE}$ to mimic the star-forming cloud core embedded in the interstellar medium, while a large domain ($r>2\,r_{\rm BE}$, interstellar medium) is left behind  from the gravitational contraction.
Thus, in our setting, the star forming cloud with a size of 5700\,AU ($5.7\times10^4$\,AU, 1740\,AU) exists in the interstellar medium with a box size of $\sim 8.7\times 10^4$\,AU ($\sim 8.7\times 10^5$\,AU, $\sim 2.7\times 10^4$\,AU) for a model with $n_{\rm c,0}=3\times10^6  \cm$ ($3\times10^4  \cm$, $3\times10^7  \cm$).
The purpose of this study is to investigate the effect of magnetic braking on disk formation.
The angular momentum is transferred by magnetic braking with the Alfv$\acute{\rm e}$n speed.
To suppress artificial reflection of Alfv$\acute{\rm e}$n waves at the computational boundary, we adopt a large simulation box to reproduce a realistic interstellar medium around the star forming cloud.
The Alfv$\acute{\rm e}$n speed in the interstellar medium ($r>2\,r_{\rm BE}$) for the model with the strongest magnetic field ($B_{\rm z,0}=176\,\mu$G)  and $\nc=3\times10^6  \cm$ is $v_{\rm A}=1.3$\,km\,s$^{-1}$.
Thus, for the Alfv$\acute{\rm e}$n wave to reach the computational boundary from the center of the cloud, it takes $3\times10^5$\,yr ($=L_1/v_{\rm A}$), which corresponds to about 30 times the free-fall timescale ($t_{\rm ff,c}=1.0\times10^4$\,yr), where the freefall timescale $t_{\rm ff,c}$ is normalized by the central density in the initial cloud.
In all models, we stopped the calculation within $t<20\,t_{\rm ff,c}$.
Thus, the Alfv$\acute{\rm e}$n wave generated at the center of the cloud (or the computational boundary) never reaches the computational boundary (or the center of the cloud).
With this setting, we can correctly treat magnetic braking in the collapsing cloud.

In the collapsing cloud core, we assume the protostar formation to occur when the number density exceeds $n > 10^{12}\cm$ at the center of the cloud.
To model the protostar, we adopt a sink around the center of the computational domain.
In the region of $r < r_{\rm sink} = 1\,$AU, gas having a number density $n > 10^{12}\cm$ is removed from the computational domain and added to the protostar as  gravity in each timestep (for details, see \citealt{machida09a,machida10a}).
This treatment of the sink makes it possible to calculate the evolution of the collapsing cloud and circumstellar disk for a longer duration.
In addition, inside the sink, the magnetic flux is removed by Ohmic dissipation, because such a region has a magnetic Reynolds $Re$ number exceeding unity $Re>1$ (for details, see \citealt{machida07}).

\section{Results}
\label{sec:results}
To understand the process of formation of a circumstellar disk in a magnetized cloud, we prepared nine different initial clouds with various masses, radii, magnetic field strengths, and angular velocities as listed in Table~\ref{table:1}.
These clouds have masses of $M_{\rm cl,0}=0.3-10\msun$ and radii of $R_{\rm cl,0}=1740-5.4\times10^4$\,AU.
The magnetic field strengths $B_0$ and angular velocities  $\Omega_0$ at the center of the initial cloud are in the range of $B_0=17.6-566\,\mu$G and $\Omega_0=(2.3-74)\times10^{-14}$\,s$^{-1}$, respectively.
In the following subsections, after we describe the disk formation for a typical model (model 1), we compare the disk formation in models with different initial settings.

\subsection{Disk Formation for a Typical Model}
\label{sec:typical}
\begin{figure}
\begin{center}
\includegraphics[width=150mm]{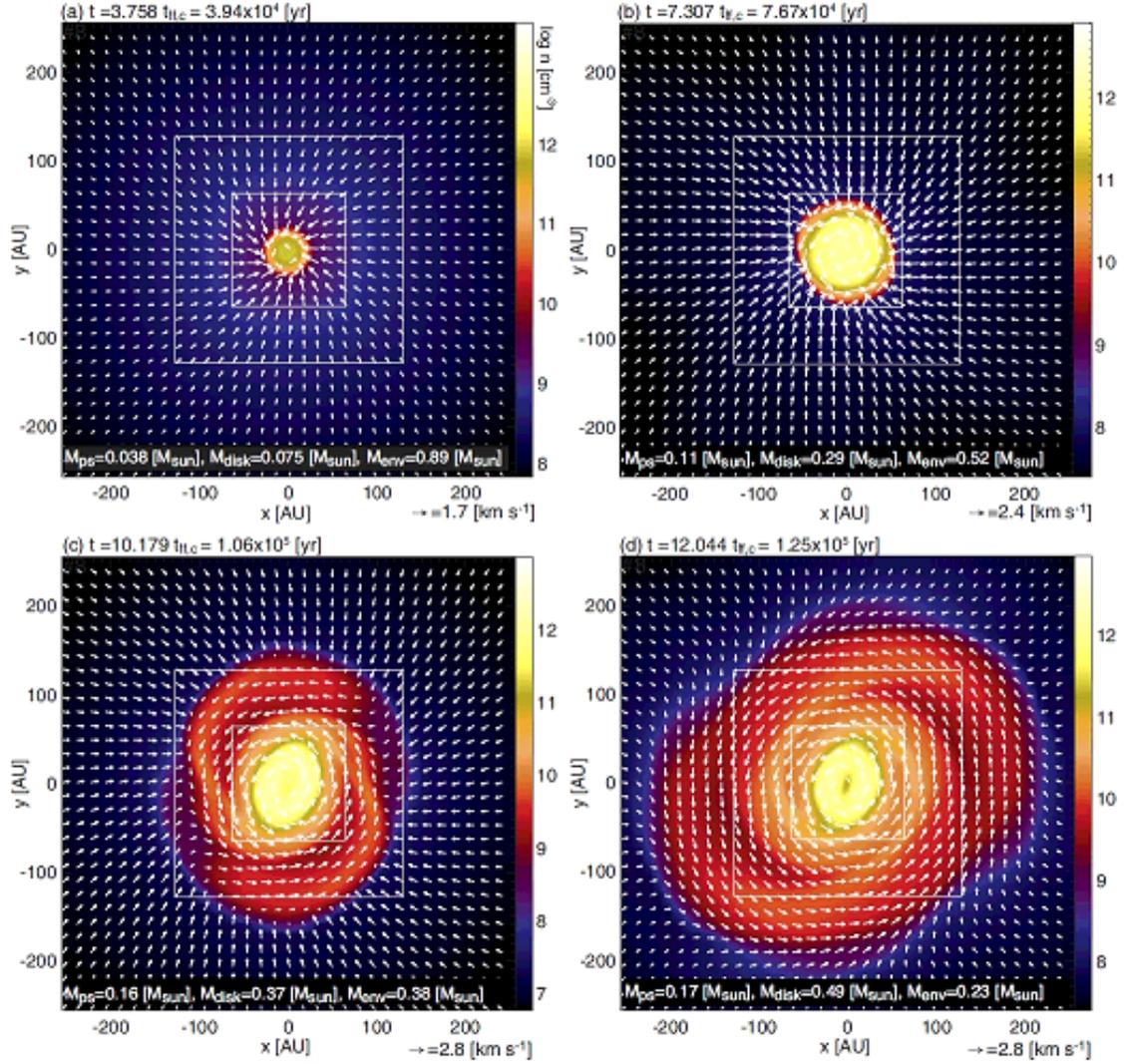}
\end{center}
\caption{
Density distribution ({\it color and contours}) and velocity vectors ({\it arrows}) on the equatorial plane at four different epochs for model 1.
The elapsed time $t$, protostellar mass $M_{\rm ps}$, circumstellar disk mass $M_{\rm disk}$, and envelope mass $M_{\rm env}$ are listed in each panel.
}
\label{fig:3}
\end{figure}

In this subsection, we describe the disk formation for model 1, which has the parameter of $\lambda=1$.
Figure~\ref{fig:3} shows the time sequence of the density distribution around the center of the collapsing cloud on the equatorial plane for model 1.
The elapsed time in units of the initial free-fall timescale $t_{\rm ff,c}$ and year after the calculation begins are described in each panel.
The protostellar, circumstellar disk, and envelope masses ($M_{\rm ps}$, $M_{\rm disk}$, and $M_{\rm env}$, respectively) are also described in each panel.
The protostellar mass $M_{\rm ps}$ is defined as the mass falling into the sink.
To determine the circumstellar disk  in the infalling envelope, we estimated the velocity ratio $R_v$ ($\equiv \vert v_r/v_\phi \vert$, where $v_r$ and $v_\phi$ are the radial and azimuthal velocity, respectively) in each cell, and specified the most distant cell having $R_v < 1$ from the center of the cloud.
Then, we defined the disk radius $r_{\rm d}$ as the distance of the cell furthest from the origin, and the disk boundary density $\rho_{\rm d,b}$ as the density of the most distant cell.
Finally, we defined the circumstellar disk as a region that has a density of $\rho > \rho_{\rm d,b}$ and estimated the circumstellar mass $M_{\rm disk}$.
%%we defined the circumstellar disk that has a density of 
\citet{machida10a} showed that the Keplerian rotating disk is well identified with this criterion.
In addition, in each timestep, we estimated the total mass $M_{\rm tot}$ that is sum of the total mass inside the gravitational sphere inside $r<2\,r_{\rm BE}$  and  the protostellar mass.
Then, we subtracted the circumstellar disk mass and protostellar mass from the total mass and derived the mass of the infalling envelope as
\begin{equation}
M_{\rm env} = M_{\rm tot} - M_{\rm disk} -M_{\rm ps}.
\end{equation}
We confirmed that the sum of the protostellar mass and the total mass is well conserved during the calculation.
Note that, in the later main accretion phase, because the mass is ejected from the collapsing cloud by the protostellar outflow, mass conservation inside $r<2\,r_{\rm BE}$ is not realized.

Figure~\ref{fig:3} shows that the circumstellar disk gradually increases its size with time.
Figure~\ref{fig:3}{\it a} indicates that the circumstellar disk has a size of $\sim10$\,AU in the early main accretion phase, as described in \citet{mellon08,mellon09}.
At this epoch, the circumstellar disk has a mass $M_{\rm disk}=0.075\msun$, which is about twice the protostellar mass ($M_{\rm ps}=0.038\msun$).
\citet{inutsuka10} showed that the circumstellar disk mass is greater than  the protostellar mass in the early main accretion phase, because the first adiabatic core, which is formed before the protostar formation with a mass of $\sim0.1-0.01\msun$, directly evolves into the circumstellar disk after the protostar formation.
In addition, at the same epoch, the mass of the infalling envelope is much larger than that of the circumstellar disk ($M_{\rm env}>10M_{\rm disk}$).

\begin{figure}
\begin{center}
\includegraphics[width=150mm]{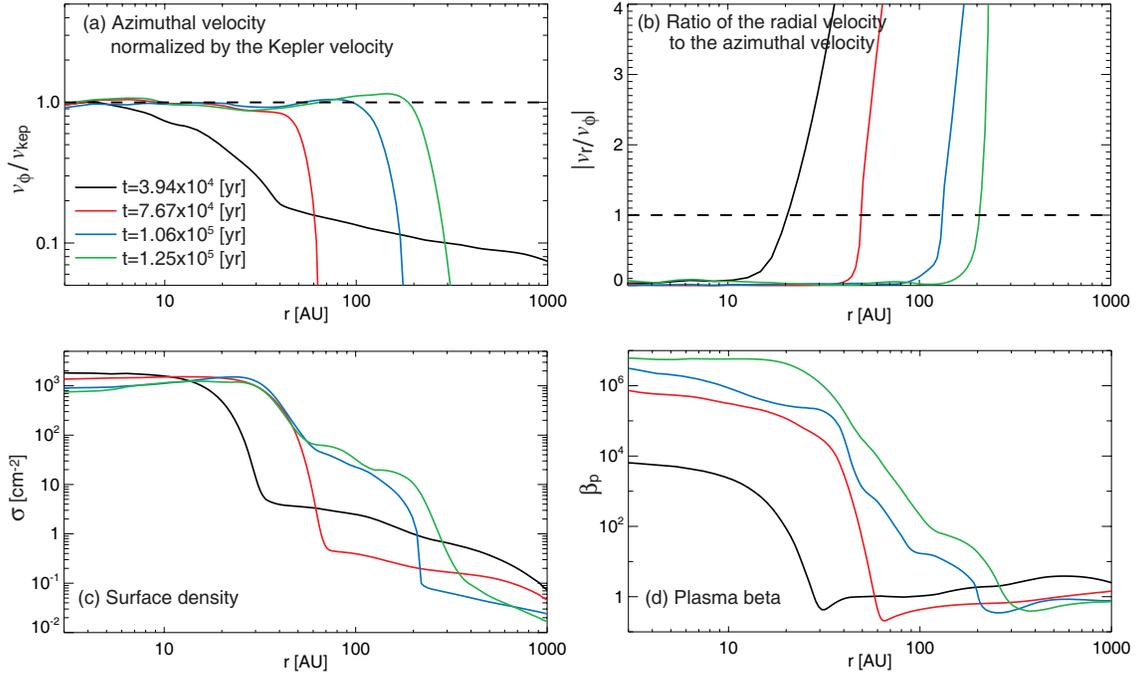}
\end{center}
\caption{
({\it a}) Azimuthal velocity normalized by the Kepler velocity, ({\it b}) ratio of the radial velocity to the azimuthal velocity, ({\it c}) surface density and ({\it d}) plasma beta $\beta_{\rm p}$ for model 1 at the same epochs as Figs.~\ref{fig:3} are plotted against the cloud radius.
Each quantity is azimuthally averaged.
}
\label{fig:4}
\end{figure}

In Figure~\ref{fig:4}, the radial distribution of the azimuthal velocity normalized by the Keplerian velocity ({\it a}), ratio of radial to azimuthal velocity ({\it b}), surface density ({\it c}) and plasma beta ({\it d}) at each epoch corresponding to each panel in Figure~\ref{fig:3} are plotted.
The azimuthal velocity and ratio of the radial to azimuthal velocity are azimuthally averaged on the equatorial plane.
The surface density is estimated as
\begin{equation}
\sigma_{\rm s} (r, \theta) = \int^{z < z_{\rm cri}} \rho(r, \theta, z)\, dz,\\
\label{eq:sigmas}
\end{equation}
where $z_{\rm cri}=50$\,AU is adopted, and azimuthally averaged.
Note that the surface density depends only a little on $z_{\rm cri}$ when $z_{\rm cri}\gg10$\,AU (and/or $z_{\rm cri}\lesssim 100$\,AU), because the disk mass is concentrated on the equatorial plane.
The plasma beta is defined in each cell on the equatorial plane as 
\begin{equation}
\beta_{\rm p} \equiv \dfrac{8\pi P}{B^2}, 
\end{equation}
where $B$ and $P$ are the magnetic flux density and thermal pressure on each cell, respectively, which are also azimuthally averaged on the equatorial plane.

At the same epoch as Figure~\ref{fig:3}{\it a} ($t=3.94\times10^4$\,yr, black lines in Fig.~\ref{fig:4}), the disk has a Keplerian velocity in the range of $r \lesssim 10$\,AU (Fig.~\ref{fig:4}{\it a}), while the disk structure extends up to $\sim20-40$\,AU (Fig.~\ref{fig:4}{\it c}).
In the range of $10\,{\rm AU} \lesssim r \lesssim 40\,{\rm AU}$, the disk has $20-80\%$ of the Keplerian velocity.
Figure~\ref{fig:4}{\it b} shows that the azimuthal velocity is greater than the radial velocity in the range of $r\lesssim 20$\,AU.
Thus, the disk is mainly supported by the centrifugal force in the range of $r\lesssim20$\,AU, while it is also supported by the thermal pressure gradient and Lorentz force in addition to the centrifugal force in the range of $20\,{\rm AU} \lesssim r \lesssim 40\,{\rm AU}$.
The black line in Figure~\ref{fig:4}{\it d} shows the rapid increase of the plasma beta at $r\sim30$\,AU.
Because the thermal pressure increases in the high-density region, the plasma beta tends to increase inside the disk.
In addition, the magnetic field dissipates by Ohmic dissipation in such a high-density region \citep{nakano02}.

Figures~\ref{fig:3} and \ref{fig:4} indicate that the circumstellar disk increases its size with time.
At $t=7.67\times10^4$\,yr (Fig.~\ref{fig:3}{\it b} and  red lines in Fig.~\ref{fig:4}), the circumstellar disk extends up to  $\sim60$\,AU and has a mass of $\sim0.29\msun$.
Figure~\ref{fig:4}{\it a} shows that the disk rotates with a Keplerian velocity in the range of $r\lesssim50-60$\,AU at this epoch.
The radial distribution of the surface density indicates that the disk extends up to $r\sim50-70$\,AU.
In addition, the plasma beta rapidly increases at $r\sim60$\,AU and reaches $\beta_{\rm p}\sim 10^6$ near the origin.
Thus, the whole region of the disk ($r\lesssim 60$\,AU) is mainly supported by the centrifugal force.
In other words, the Keplerian rotating disk extends up to $\sim60$\,AU at this epoch.

The circumstellar disk mass becomes comparable to the mass of the infalling envelope at $t\simeq10^5$\,yr, which corresponds to $t\simeq10\,t_{\rm ff,c}$ (Fig.~\ref{fig:3}{\it c} and  blue lines in Fig.~\ref{fig:4}).
At the initial state, the density at the cloud boundary (i.e., $r=2\,r_{\rm BE}$) is 82 times lower than that at the center of the cloud ($r=0$).
Thus, the freefall timescale at the cloud boundary $t_{\rm ff,b}$ is about nine times longer than that at center of the cloud $t_{\rm ff,c}$ (i.e., $t_{\rm ff,b}= \sqrt{82} f_{\rm ff,c} = 9.1f_{\rm ff,c}$). 
Thus, we can expect that a large fraction of the initial cloud mass has already fallen onto the central region at $t\gg9f_{\rm ff,c}$.
Note that the cloud rotation and magnetic field slightly slows the cloud collapse and accretion onto the disk (e.g., \citealt{machida05a,machida06a}).
Figure~\ref{fig:3}{\it c} shows that the size of the circumstellar disk exceeds 100\,AU and the spiral structure appears in the disk at $t=1.06\times10^5\,{\rm yr}=10.2\,t_{\rm ff,c}$.
The blue lines in Figure~\ref{fig:4} also indicate that the Keplerian rotating disk extends up to $r\gtrsim100$\,AU at this epoch.

At $t=1.25\times10^5\,{\rm yr}=12.04\,t_{\rm ff,c}$ (Fig.~\ref{fig:3}{\it d} and green lines in Fig.~\ref{fig:4}), the circumstellar disk mass ($M_{\rm disk}=0.49\msun$) is much greater than the mass of the infalling envelope ($M_{\rm env}=0.23\msun$).
Thus, as described in \S\ref{sec:intro}, it is expected that magnetic braking is not effective at this epoch (for details, see \S\ref{sec:diss}).
The green lines in Figure~\ref{fig:4} indicates that the circumstellar disk with a Keplerian velocity has a size of $\sim200$\,AU (see Fig.~\ref{fig:4}{\it c}) at this epoch.
Thus, the circumstellar disk exceeding $100$\,AU can form even in a strongly magnetized cloud core.

\begin{figure}
\begin{center}
\includegraphics[width=150mm]{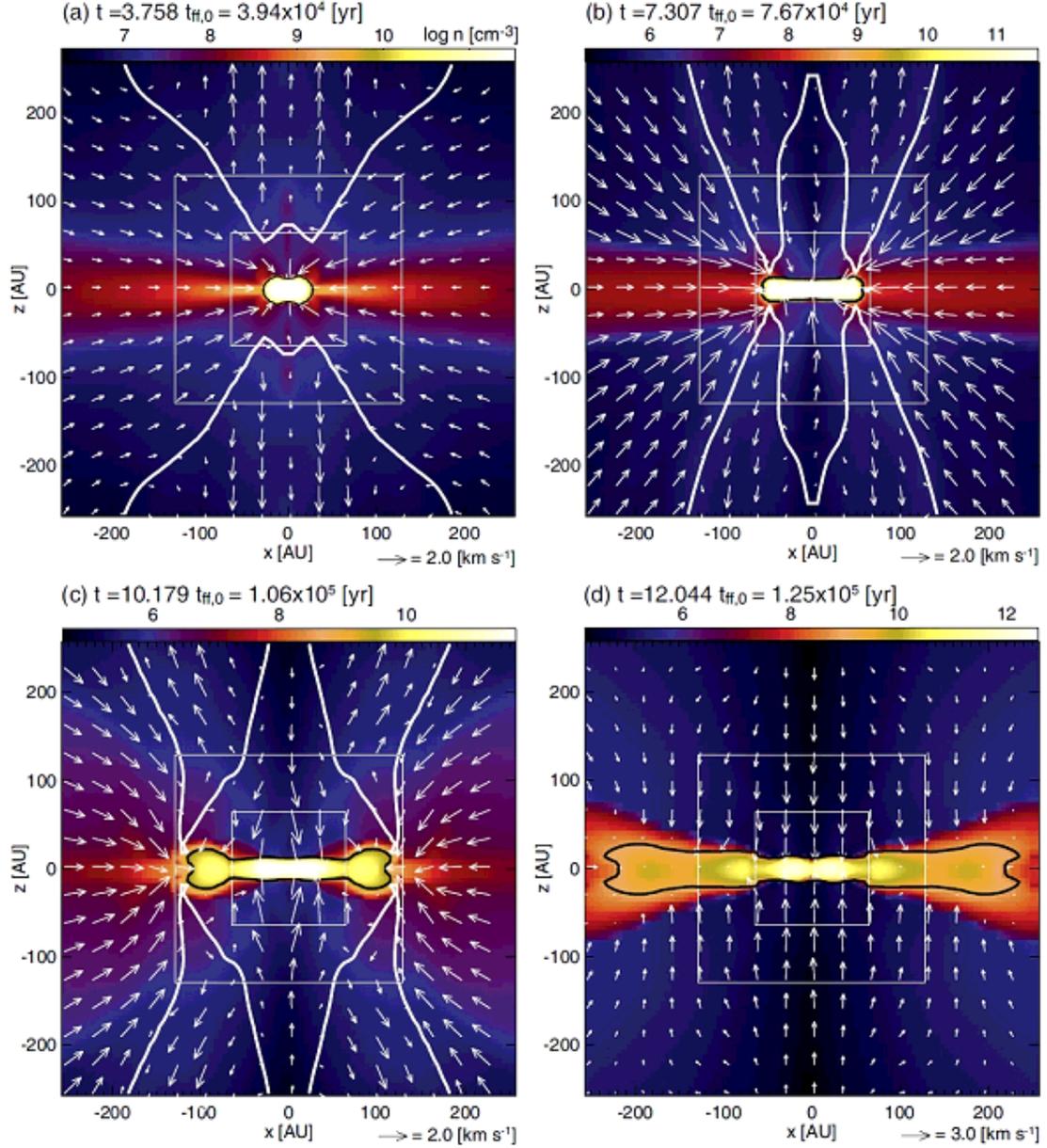}
\end{center}
\caption{
Same as Figure~\ref{fig:3} but on the $y=0$ plane.
In each panel, the black line corresponds to the circumstellar disk.
The boundary between outflowing ($v_r>0$) and inflowing ($v_r<0$) gas is described by the white line; inside this boundary, the gas is outflowing from the circumstellar disk.
}
\label{fig:5}
\end{figure}

Figure~\ref{fig:5} shows the time sequence of the density distribution for model 1 at the same epochs as in Figure~\ref{fig:3} but for the $y=0$ plane.
The outflow is driven near the circumstellar disk, as seen in Figure~\ref{fig:5}{\it a}-{\it c}.
This type of outflow was seen in many previous studies (e.g., \citealt{tomisaka02,machida04,machida05b,machida06b,machida08b,hennebelle08,duffin09,tomida10}).
In each panel, the Keplerian rotation disk is indicated by the black circle, inside which the gas rotates with a (nearly) Keplerian speed.
The figure shows that the high-density Keplerian rotating disk (black line) is enclosed by a low-density pseudo disk.
The pseudo-disk is formed by both the cloud rotation and magnetic field in the gas collapsing phase \citep{machida05a}.
However, the pseudo-disk gradually disappears in the main accretion phase, because its rotation speed does not reach the Keplerian velocity.
Figures~\ref{fig:5}{\it a}-{\it c} show that the driving point of the outflow radially expands with time, because the Keplerian rotating disk, which drives the outflow, extends in the radial direction with time.
The outflow disappears in $\sim1.2\times10^5$\,yr after the cloud collapse begins.
In Figure~\ref{fig:5}{\it d}, there is a sufficiently thin Keplerian rotation disk with a size of $\sim200$\,AU, while no outflow is driven by the circumstellar disk at this epoch.
This is because by this epoch almost all the cloud mass already has fallen onto either the circumstellar disk or the protostar.
The outflow is powered by the gas accreting energy onto the circumstellar disk.

\begin{figure}
\begin{center}
\includegraphics[width=150mm]{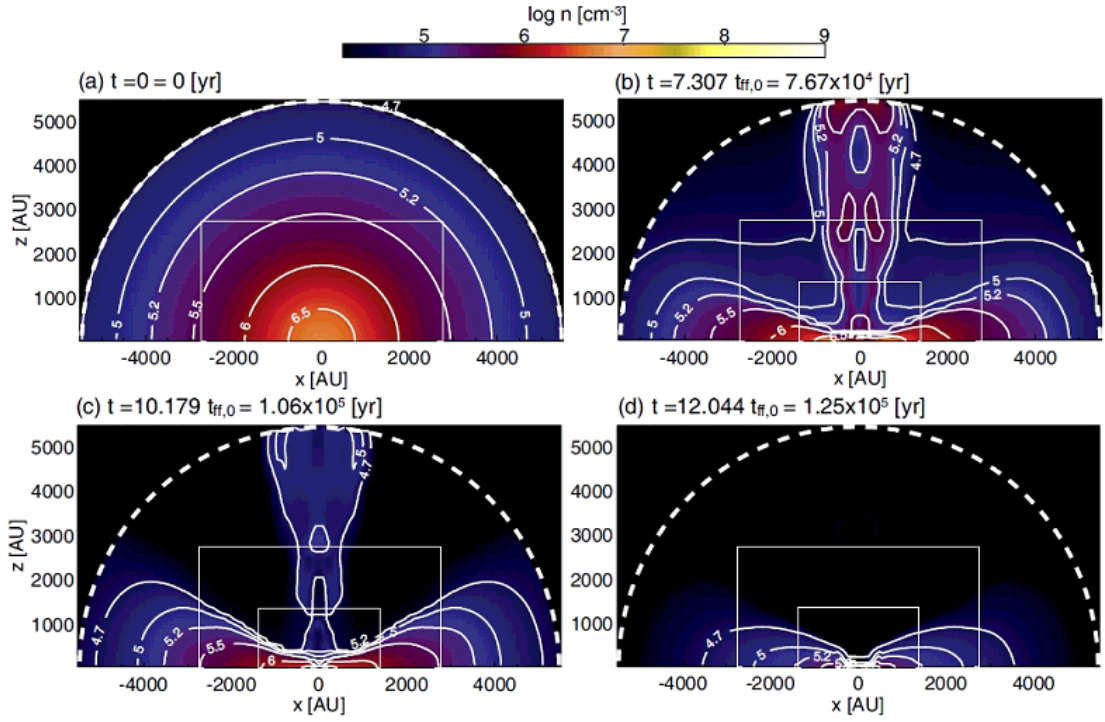}
\end{center}
\caption{
Density distribution ({\it color and contours}) for model 1 in the region of $z>0$ on the $y=0$ plane at the same epochs as Figs.~\ref{fig:3} and \ref{fig:5} except for panel {\it a} (initial state).
%%The spatial scale is different from that in Figs.~\ref{fig:3} and \ref{fig:4}.
The white broken line is the boundary between the gravitational sphere and interstellar medium; gravity (gas self-gravity and the gravity of the protostar) is imposed inside this boundary.
}
\label{fig:6}
\end{figure}

To investigate the evolution of the infalling envelope, the density distribution on the $y=0$ plane at the same epochs as in Figures~\ref{fig:3} and \ref{fig:5} but for the initial cloud scale ($\sim10^5$\,AU) is plotted in Figure~\ref{fig:6}.
In each panel, the boundary between the gravitational sphere and static interstellar medium ($r=2\,r_{\rm BE}$) is plotted by a white broken line; gas only inside the boundary can fall toward the center of the cloud.
The figure shows that the cloud density at a larger radius gradually decreases with time, and the gas is concentrated onto the equatorial plane.
The gas is also distributed along the $z$-axis in Figures~\ref{fig:6}{\it b} and {\it c}.
This mass distribution is caused by the mass ejection from the circumstellar disk by the protostellar outflow.
Figure~\ref{fig:6}{\it d} shows that gas in the infalling envelope is depleted and the mass ejection by the protostellar outflow has already stopped.
It is natural that the elapsed time at Figure~\ref{fig:6}{\it d} ($t=12\,t_{\rm ff,c}$) is longer than the freefall timescale ($t_{\rm ff,b} = 9.1\,t_{\rm ff,c}$) at the cloud boundary, and thus the gas located near the cloud boundary at the initial state has already fallen onto the circumstellar disk by this epoch.
It is expected that, at this epoch,  magnetic braking becomes ineffective because the (less massive) infalling envelope cannot brake the circumstellar disk.
Therefore, a large-size Keplerian rotating disk can appear at this epoch.

Note that we may overestimate the mass of the infalling envelope, because we defined it  as the mass of the whole cloud  except for the Keplerian rotating disk.
Thus, the infalling envelope includes the pseudo disk in our definition.
The infalling envelope has a mass of $M_{\rm env}=0.23\msun$ (i.e., $23\%$ of the initial cloud mass) at $t=1.25\times10^5$\,yr, as described in Figure~\ref{fig:3}{\it d}.
In reality, however, a very small amount of the infalling gas remains in the region above the circumstellar disk as seen in Figure~\ref{fig:6}{\it d}.

\begin{figure}
\begin{center}
\includegraphics[width=150mm]{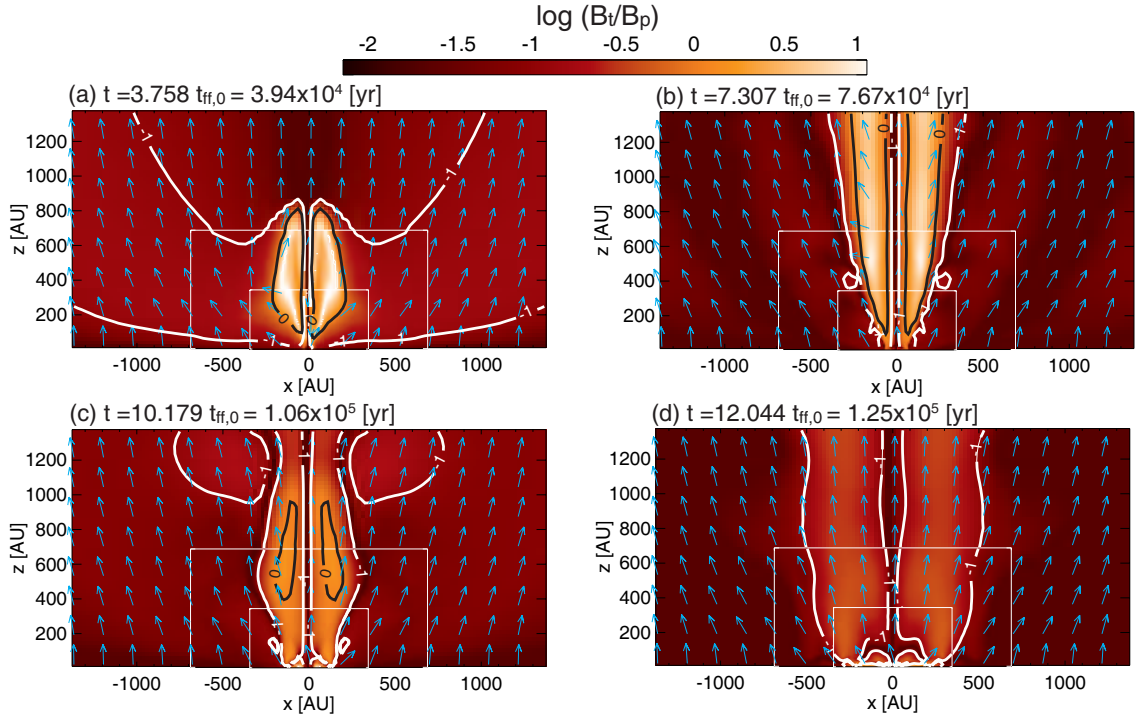}
\end{center}
\caption{
Ratio of the toroidal field to the poloidal field ({\it color and contours}) and magnetic field ({\it arrows}) for model 1 in the region of $z>0$ on the $y=0$ plane at the same epochs as Figures~\ref{fig:3}, \ref{fig:5} and \ref{fig:6}.
The toroidal component ($B_{\rm t}$) is greater than the poloidal component ($B_{\rm p}$) inside the black line.
The white line corresponds to the contour of $B_{\rm t}=0.1 B_{\rm p}$.
The arrows indicate the magnetic field direction  on the $y=0$ plane.
}
\label{fig:7}
\end{figure}

The circumstellar disk is connected to the infalling envelope through the magnetic field lines.
Thus, a slowly rotating massive infalling envelope can brake a rapidly rotating less massive disk through the magnetic field line.
Thus, the magnetic field lines are strongly twisted when the infalling envelope is more massive than the circumstellar disk or when magnetic braking is effective.
The difference between the rotation speeds of the infalling envelope and the circumstellar disk generates a strong toroidal field. 
Instead, when the infalling envelope is much less massive than the circumstellar disk,  the infalling envelope cannot brake the circumstellar disk, and a strong toroidal field is not generated.
In such a case, it is expected that the magnetic field has a nearly force free configuration.
Figure~\ref{fig:7} shows the ratio of the toroidal field to the poloidal field ($B_t/B_p$) at the same epochs as in Figure~\ref{fig:3}.
The toroidal field is greater than the poloidal field (toroidal dominated region; $B_r>B_p$) inside the black line.
Figure~\ref{fig:7}{\it a} shows that the toroidal dominated region is distributed along the $z$-axis in the early main accretion phase, in which the torsional Alfv$\acute{\rm e}$n wave generated by the rotation of the circumstellar disk is propagated along the $z$-axis.

In Figure~\ref{fig:7}{\it b}, the toroidal field is much greater than the poloidal field in the region above the circumstellar disk (i.e., in the region of $ r_c  \lesssim 100-200$\,AU, where $r_c$ is the cylindrical radius).
This indicates that magnetic braking is effective and the angular momentum of the circumstellar disk is transferred into the infalling envelope.
At this epoch, strong mass ejection from the circumstellar disk also takes place as seen in Figure~\ref{fig:6}{\it b}, resulting from the generation of the strong toroidal field.
The toroidal field weakens for $t\gtrsim 10\, t_{\rm ff,c}$ as seen in Figure~\ref{fig:7}{\it c}.
In Figure~\ref{fig:7}{\it d}, the poloidal field dominates the toroidal field in the whole region of the infalling envelope. 
By this epoch, a large fraction of the mass in the initial cloud has already fallen onto either the circumstellar disk or the protostar.
At this epoch, the mass of the infalling envelope is much less than the circumstellar disk mass (Fig.~\ref{fig:6}{\it d}).
Thus, it is expected that, at this epoch, the magnetic field is rarely twisted and magnetic braking is not so effective.

\begin{figure}
\begin{center}
\includegraphics[width=150mm]{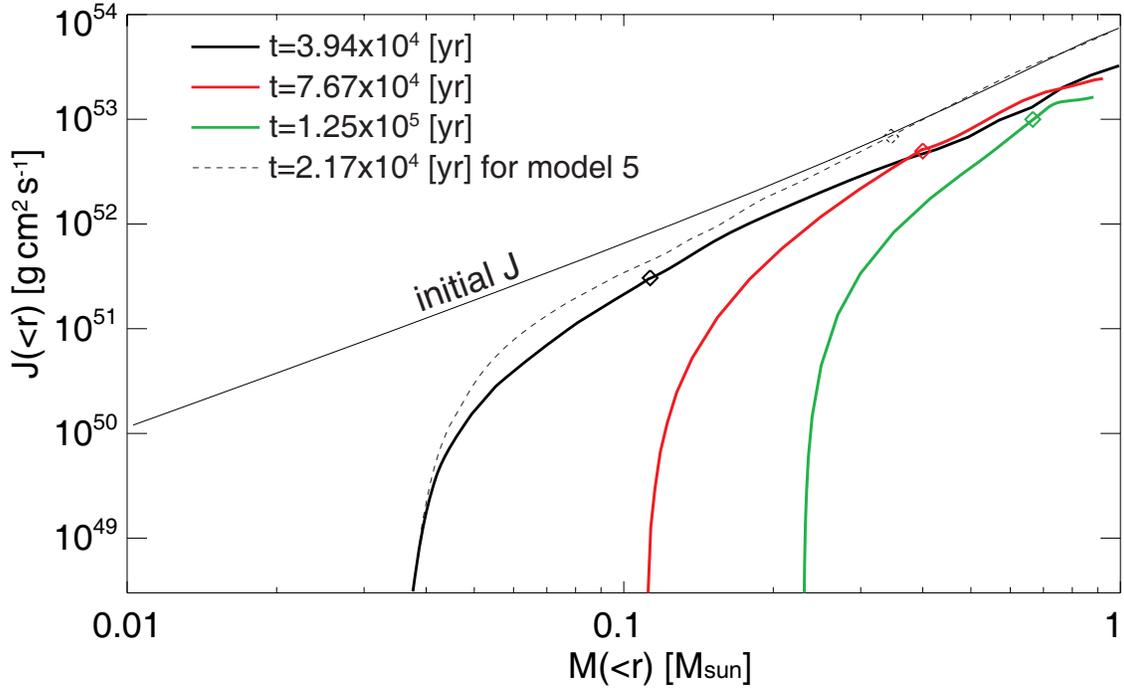}
\end{center}
\caption{
Distribution of the angular momentum for model 1 at the same epochs as in Figure~\ref{fig:3} against the cloud mass.
The dashed line is the distribution of the angular momentum for model 5 at $t=1.67\times10^4$\,yr.
The initial distribution of the angular momentum is plotted by a thin solid line.
The angular momentum and mass are integrated in the region of $<r$.
The sum of the protostellar mass $M_{\rm ps}$ and circumstellar disk mass $M_{\rm disk}$ of each epoch is plotted by the diamond symbol ($\diamond$) on each line.
The region on the left-hand side of the diamond symbol corresponds to the circumstellar disk region.
}
\label{fig:8}
\end{figure}

Figure~\ref{fig:8} shows the radial distribution of the angular momentum for model 1 at the same epochs as in Figure~\ref{fig:3}  against the radially integrated  mass.
For reference, the distribution of the angular momentum for unmagnetized model (model 5) at almost the same epoch as Figure~\ref{fig:3}{\it a} is also plotted in this figure.
At each radius, the angular velocity and mass are integrated in the gravitationally collapsing cloud ($r<2\,r_{\rm BE}$) as
\begin{equation}
J(r) = \int^r_0 \rho\, r_c\, v_{\phi}\, dv,
\end{equation}
and
\begin{equation}
M(r) = M_{\rm ps} + \int^r_0 \rho \, dv,
\end{equation}
where the protostellar mass $M_{\rm ps}$ falling into the sink is added to estimate the mass.
Note that, in Figure~\ref{fig:8}, the cloud mass does not reach $M=1\msun$ except for unmagnetized model (dotted line) because some fraction of the cloud mass ($\sim10$\%) is ejected from the collapsing cloud (or from the region of $r\le2\,r_{\rm BE}$) by the protostellar outflow. 
The figure indicates that the angular momentum of the whole cloud ($M\sim1\msun$) decreases gradually with time for model 1, while it is well conserved in the whole cloud for model 5.
For model 1, the angular momentum of the cloud can be transferred into the interstellar medium by magnetic braking.
However, the rate of reduction of the angular momentum is not very large during the calculation.
The initial cloud has an angular momentum of $J=5.5\times10^{53}\,\jj$, while the collapsing cloud has $J=2.1\times10^{53}\,\jj$ as the whole at the end of the calculation.
Thus, about half of the angular momentum in the initial cloud is transferred into the interstellar medium.
This is natural because the timescale of magnetic braking by the interstellar medium is longer than our calculation time (discussed in \S\ref{sec:diss}).
%%Note that the magnetic braking timescale due to the interstellar medium 
%%The timescale of the magnetic braking from the interstellar medium is discussed in \S\ref{sec:diss}.
In addition, because the protostellar outflow driven by the circumstellar disk propagates into the interstellar medium with a certain amount of angular momentum, the protostellar outflow also transfers the angular momentum into the interstellar medium.

Figure~\ref{fig:8} also indicates that the angular momentum inside the disk (i.e., the region on the left side of the diamond symbol) is considerably smaller than that at the initial state.
This difference of the angular momentum between the initial state and collapsing cloud clearly indicates that the angular momentum is effectively transferred outward in the collapsing cloud.
In the early main accretion phase, the angular momentum is expected to be transferred by magnetic effects such as magnetic braking and protostellar outflow, since the circumstellar disk has an axisymmetric structure and remains clear axisymmetricity (Figs.~\ref{fig:3}{\it a} and {\it b}), and hence, gravitational torque seems to be less important in magnetized models.
%%has  
In the later main accretion phase, since non axisymmetric patterns appear in the circumstellar disk (Figs.~\ref{fig:3}{\it c} and {\it d}), the angular momentum is expected to be transferred also by gravitational torque due to the non axisymmetric perturbation, in addition to the magnetic effects.

\subsection{Disk Formation in Clouds with Different Parameters}
In \S\ref{sec:typical}, we investigated the disk formation for a typical model (model 1).
However, it is expected that the formation process of the circumstellar disk and its size and mass depend on the initial cloud properties such as the magnetic field strength, rotation rate, and cloud mass (or size).
In this subsection, we show the disk formation in clouds with different parameters.

\subsubsection{Different Magnetic Field Strengths}
\label{sec:mags}
\begin{figure}
\begin{center}
\includegraphics[width=150mm]{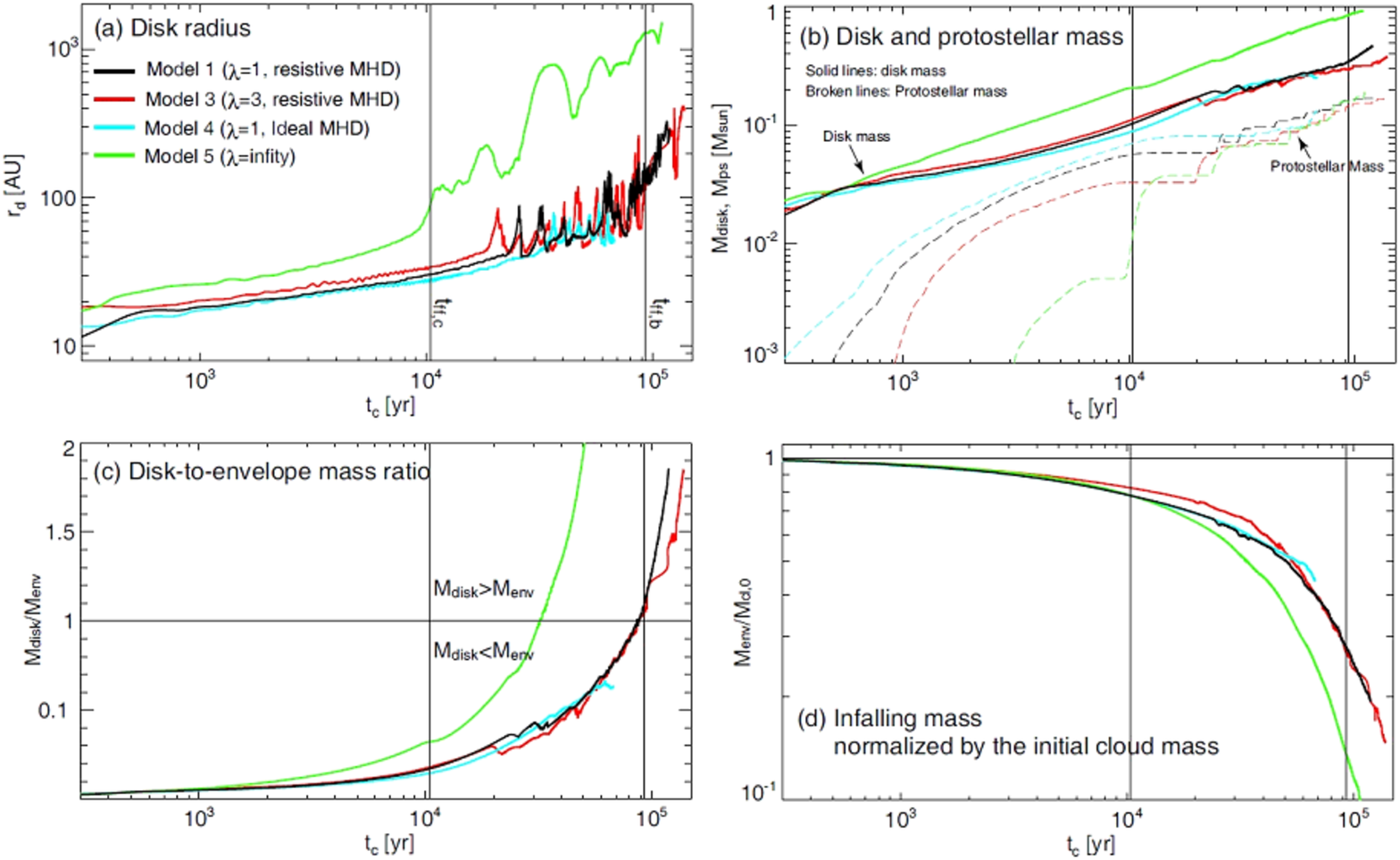}
\end{center}
\caption{
Disk radii ({\it a}), disk-to-envelope mass ratios ({\it b}), circumstellar disk and protostellar masses ({\it c}) and masses of the infalling envelope ({\it d}) for models 1, 3, 4 and 5 against the elapsed time after the circumstellar disk formation.
The freefall timescale at the center of the cloud $t_{\rm ff,c}$ and at the cloud boundary $t_{\rm ff,b}$ are plotted in each panel.
}
\label{fig:9}
\end{figure}

To investigate the disk formation in clouds with initially different magnetic field strengths, we plotted the disk radii ({\it a}), disk and protostellar masses ({\it b}), disk-to-envelope mass ratios ({\it c}) and the masses of the infalling envelope ({\it d}) for models 1, 3, 4 and 5 against the elapsed time $t_{\rm c}$ in Figure~\ref{fig:9}.
The disk is defined as the region of $R_v<1$ (see, \S\ref{sec:typical}).
Thus, in our definition, the disk (or the circumstellar disk) has a (nearly) Keplerian rotation.
Note that the elapsed time $\tc$, which is defined as the elapsed time after the circumstellar disk formation, differs from $t$, which is defined as the elapsed time after the cloud collapse begins (or the calculation begins).
They are related  as
\begin{equation}
t = t_0 + \tc,
\end{equation}
where $t_0$ is the time from the beginning of the calculation until the disk formation, and is listed in Table~\ref{table:2}.
In this subsection, we use $\tc$ in place of $t$ to compare the disk evolution from the time just after the disk formation for different models.
For models in Figure~\ref{fig:9}, model 3 has a three times larger initial mass-to-flux ratio ($\lambda=3$) than model 1.
Model 4 has the same mass-to-flux ratio ($\lambda=1$) as model 1, but the cloud evolution is calculated with the ideal MHD approximation.
The initially unmagnetized cloud ($\lambda=\infty$) is adopted for model 5.
The elapsed time from the beginning of the calculation ($t$), disk radius ($r_d$), mass of the circumstellar disk ($M_{\rm disk}$) in each model at the end of the calculation are also listed in Table~\ref{table:2}.

Figure~\ref{fig:9}{\it a} indicates that the disk radius for an unmagnetized cloud (model 5) is much larger than the radii for magnetized clouds (models 1, 3, 4).
In other words, in a magnetized cloud, the circumstellar disk is a relatively smaller that that in an unmagnetized cloud.
This indicates that the magnetic field suppresses the disk formation, as already described in past studies \citep{mellon08,mellon09,duffin09,hennebelle09}.
At $\tc\sim10^5$\,yr, which almost corresponds to the freefall timescale at the cloud boundary, the circumstellar disk for model 5 reaches $\sim1200$\,AU, which corresponds to about 20\% of the centrifugal radius of the whole cloud (see, Table~\ref{table:1}).
For model 5, the increase and decrease of the disk radius for $t>t_{\rm ff,c}\simeq10^4$\,yr is attributed to the development of the spiral structure.
As seen in Figure~\ref{fig:9}{\it b}, because the circumstellar disk mass is greater than the protostellar mass in the main accretion phase, the disk is unstable against its gravity and frequently shows a spiral structure \citep{machida10a}.
The spiral structure effectively transfers the angular momentum outward and  promotes a rapid mass accretion from the circumstellar disk onto the protostar.
After the rapid mass accretion, the mass and size of circumstellar disk temporally decreases and the spiral structure weakens.
For model 5, the circumstellar disk mass is greater than the mass of the infalling envelope for $\tc\gtrsim3\times10^4$\,yr (Fig.~\ref{fig:9}{\it c}).
In addition, over 90\% of the initial cloud mass already falls either onto the circumstellar disk or protostar at $\tc\simeq10^5$\,yr (Fig.~\ref{fig:9}{\it d}).
Thus, it is expected that the main accretion phase almost ends by this epoch, and the circumstellar disk grows only a little by gas accretion after this epoch.

On the other hand, at $\tc \sim 10^5$\,yr, the disk radii in magnetized clouds are about $100-200$\,AU, which is considerably smaller than that of the unmagnetized cloud ($\sim1200$\,AU).
The large difference of disk radii between magnetized and unmagnetized models indicates that a large fraction of the angular momentum is transferred  by magnetic effects.
The difference of disk radii is caused by the different efficiencies of the angular momentum transfer among models.

In Figure~\ref{fig:9}{\it a}, resistive MHD models (models 1 and 3) show a similar trend for the evolution of the disk radius, in which model 3 has a three times larger mass-to-flux ratio than model 1 at the initial state.
For these models, the disk radii gradually increases with time for $t<(2-3)\times10^4$\,yr.
Reflecting the difference in the initial  mass-to-flux ratio, at the same epoch during the early main accretion phase [$t<(2-3)\times10^4$\,yr], model 3 which has an initially weaker magnetic field,  has a slightly larger disk than model 1.
Then, for $t>(2-3)\times10^4$\,yr, the disk radii for these models increase with strong oscillation, which is caused by the spiral structure appearing in the circumstellar disk.
Figure~\ref{fig:9}{\it b} shows that the circumstellar disk is more massive than the protostar by the end of the calculation even in magnetized clouds.
In addition, the protostellar mass increases intermittently for $\tc\gtrsim2\times10^4$\,yr.
This is because the spiral structure, which appears in the circumstellar disk for $\tc\gtrsim10^4$\,yr (see, Fig.~\ref{fig:3}{\it c} and {\it d}), unsteadily transfers the angular momentum outward and the mass accretion onto the protostar is temporarily amplified.
Thus, during this phase ($t\gtrsim10^4$\,yr), the angular momentum is transferred not only by the magnetic effects but also by the non axisymmetric perturbation in the circumstellar disk.

Finally, in resistive models (models 1 and 3), the disk radius rapidly increases for $\tc>(0.8-1)\times10^5$\,yr and reaches $\sim300-400$\,AU at the end of the calculation.
Figure~\ref{fig:9}{\it c} shows that the circumstellar disk mass is greater than the mass of the infalling envelope for $t\gtrsim t_{\rm ff,b}\simeq10^5$\,yr. 
Note that since the infalling envelope includes the pseudo disk in our definition as described in \S\ref{sec:typical}, we somewhat overestimate the actual infalling envelope in Figure~\ref{fig:9}{\it c}.
After this epoch ($t\gtrsim 10^5$\,yr), it is expected that magnetic braking becomes ineffective, because the less massive infalling envelope can no longer brake the circumstellar disk.
Thus, the circumstellar disk can rapidly grow without effective angular momentum transfer by the magnetic effects.
As shown in Figure~\ref{fig:9}{\it d}, for models 1 and 3, over 80\% of the initial cloud mass already falls either onto the circumstellar disk or the protostar by the end of the calculation.
Note that a part of the initial cloud mass is ejected from the gravitationally collapsing cloud by the protostellar outflow as shown in Figure~\ref{fig:6}.
At the end of the calculation, the circumstellar disk has a size of 290 AU for model 1 and 370 AU for model 3.
Therefore, although the disk size in resistive MHD models is considerably smaller than that in the unmagnetized model, the Keplerian rotation  disk with a size of $>100$\,AU can form even in strongly magnetized cloud.

\subsubsection{Resistive vs. Ideal MHD Models}
Figure~\ref{fig:9}{\it a} shows that the disk radius in the ideal MHD model (model 4) is somewhat smaller than that in the resistive MHD model (model 1) with the same parameters for the initial cloud.
At end of the calculation ($t\simeq2\times10^5$\,yr), the circumstellar disk has a size of $290$\,AU for model 1 and $120$\,AU for model 4.
Magnetic braking is more effective in the ideal MHD model than in the resistive MHD model, because the magnetic field largely dissipates inside the circumstellar disk in the resistive MHD model.
Thus, it is considered that the difference of the disk size between the ideal and resistive MHD models is caused by the difference in the strengths of the magnetic field in the circumstellar disk.

\begin{figure}
\begin{center}
\includegraphics[width=140mm]{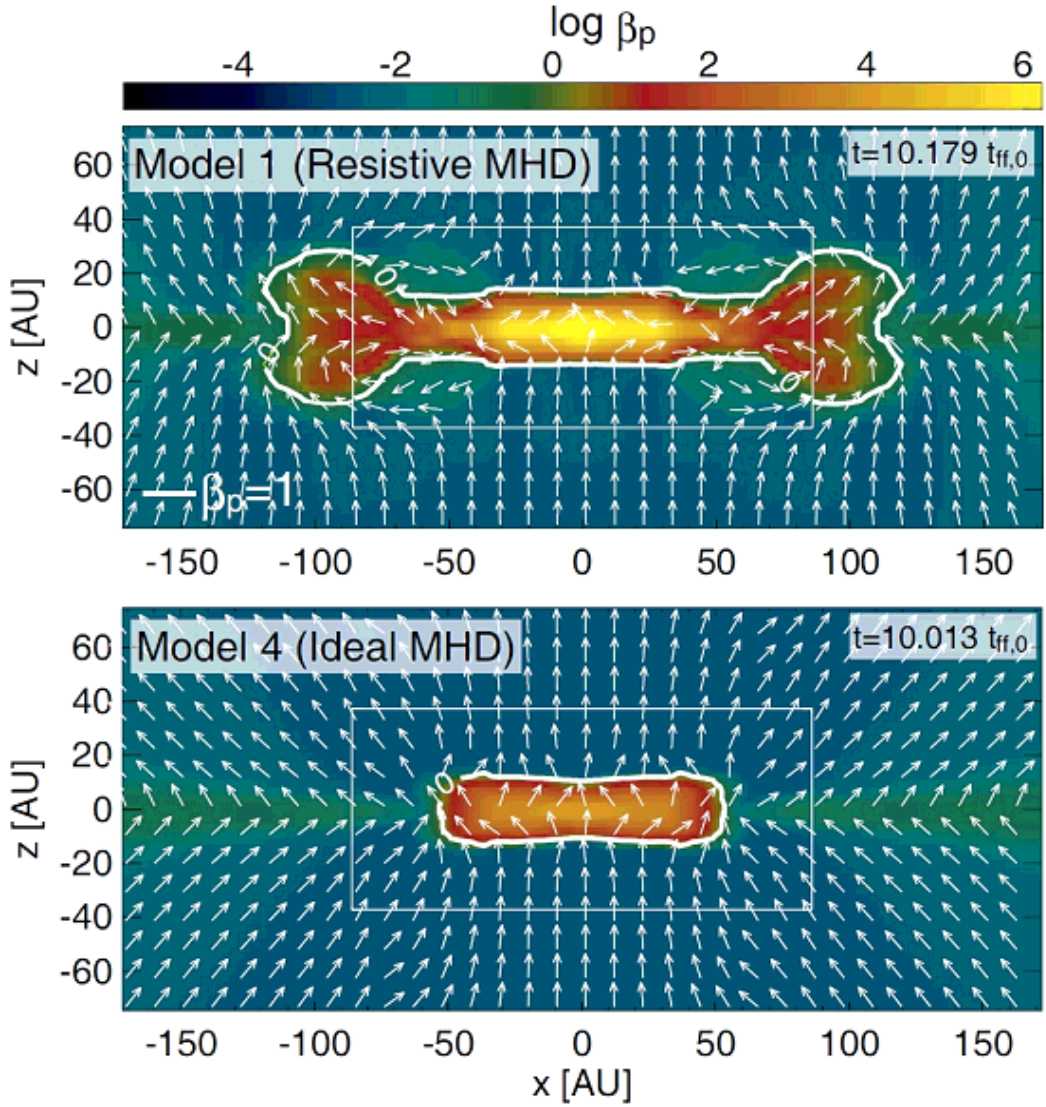}
\end{center}
\caption{
Plasma beta ({\it color}) on the $y=0$ plane for the resistive (model 1; upper panel) and ideal (model 4; lower panel) MHD models.
The thick white line is the contour of $\beta_p = 1$. 
The arrows indicate the magnetic field direction on the $y=0$ plane.
}
\label{fig:10}
\end{figure}

Figure~\ref{fig:10} shows the plasma beta for models 1 (upper panel) and 4 (lower panel) on the $y=0$ plane at almost the same epoch after the calculation begins.
The contour of $\beta_p=1$ is shown by a white thick line, outside of which the magnetic energy is greater than the thermal energy.
The circumstellar disk exists in the region  of $\beta_p\gtrsim1$ (i.e., inside the white line).
Because the circumstellar disk has a high density (or large thermal pressure), the thermal energy is greater than the magnetic energy (i.e., $\beta_p>1$) inside the circumstellar disk.
Figure~\ref{fig:10} shows that, for model 1,  the plasma beta is extremely high as $\beta_p\gtrsim 10^5 $ in the region of $\vert r \vert \lesssim 50$\,AU inside the circumstellar disk, because this region has a density of $n\sim10^{12}\cm$ (see Fig.~\ref{fig:5}{\it c}) and the Ohmic dissipation is effective.
Figure~\ref{fig:10} lower panel shows that the circumstellar disk for the ideal MHD model is smaller than that for the resistive MHD model.
In the circumstellar disk, the plasma beta for the ideal MHD model is about two or three orders of magnitude smaller than that for the resistive MHD model.
Despite the large difference in the magnetic energy in the circumstellar disk, the difference in the final size of the circumstellar disk is not very large: the disk has a size of $290$\,AU for model 1, and $120$\,AU for model 4.
In addition, the evolution of the masses of circumstellar disk, protostar, and infalling envelope for the ideal MHD model is almost the same as that for the resistive MHD model, as shown in Figure~\ref{fig:9}{\it b} and {\it c}.
Thus,  even in the ideal MHD model,  magnetic braking becomes ineffective as the mass of the infalling envelope is depleted, and a circumstellar disk with a size of $\sim100$\,AU can form.

\subsubsection{Effect of Initial Cloud Rotation}
\begin{figure}
\begin{center}
\includegraphics[width=140mm]{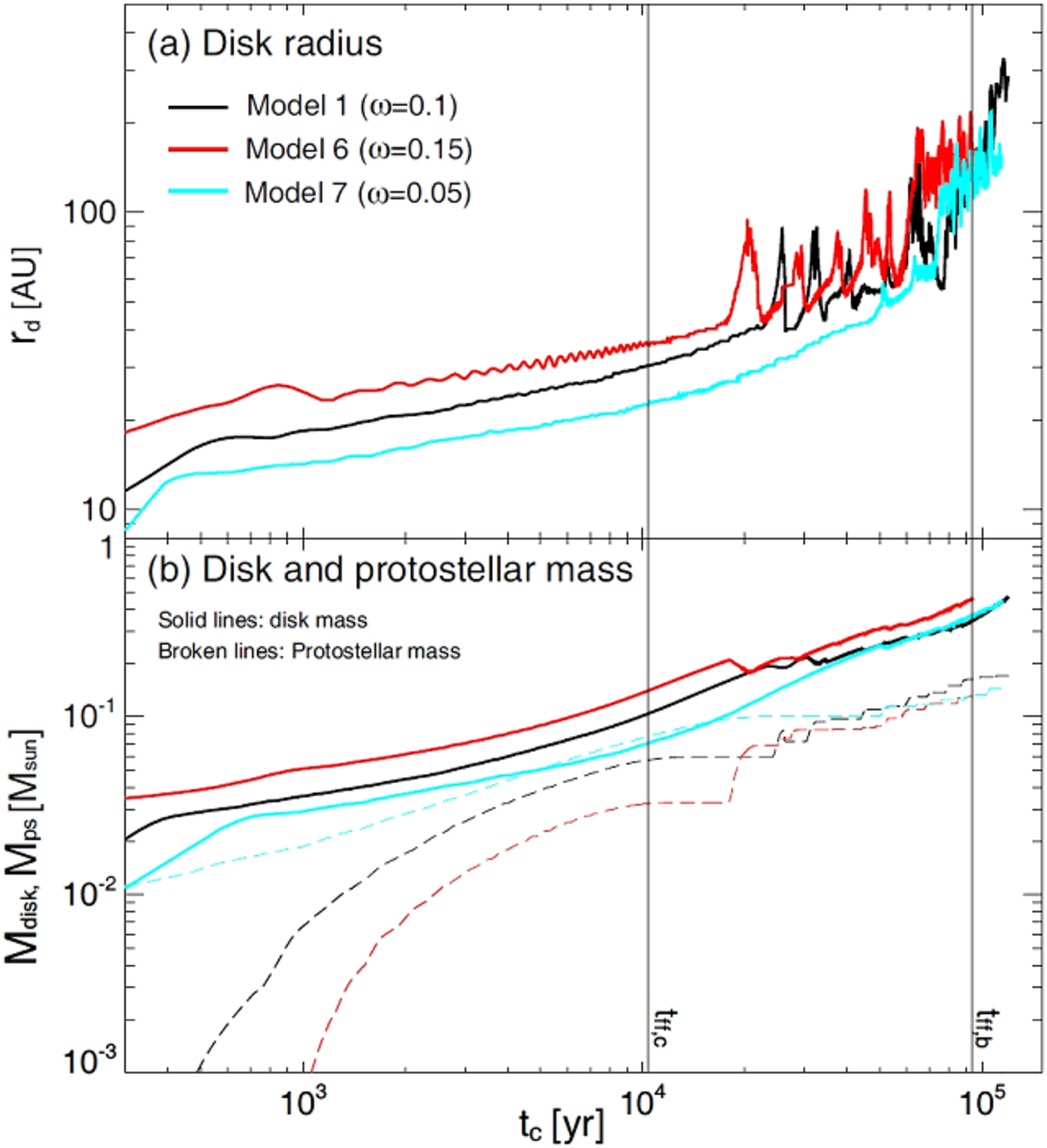}
\end{center}
\caption{
Disk radii (upper panel) and disk and protostellar masses (lower panel) for models 1, 6 and 7 against the elapsed time after the circumstellar disk formation.
The freefall timescale at the center of the cloud $t_{\rm ff,c}$ and at the cloud boundary $t_{\rm ff,b}$ are plotted in each panel.
}
\label{fig:11}
\end{figure}

To investigate the effect of the initial cloud rotation on the circumstellar disk formation, the disk radii and circumstellar disk and protostellar masses for models 1, 6 and 7 are plotted in Figure~\ref{fig:11}.
Model 6 has three times the initial angular velocity as model 1, while model 7 has one third the initial angular velocity as model 1.

\begin{figure}
\begin{center}
\includegraphics[width=140mm]{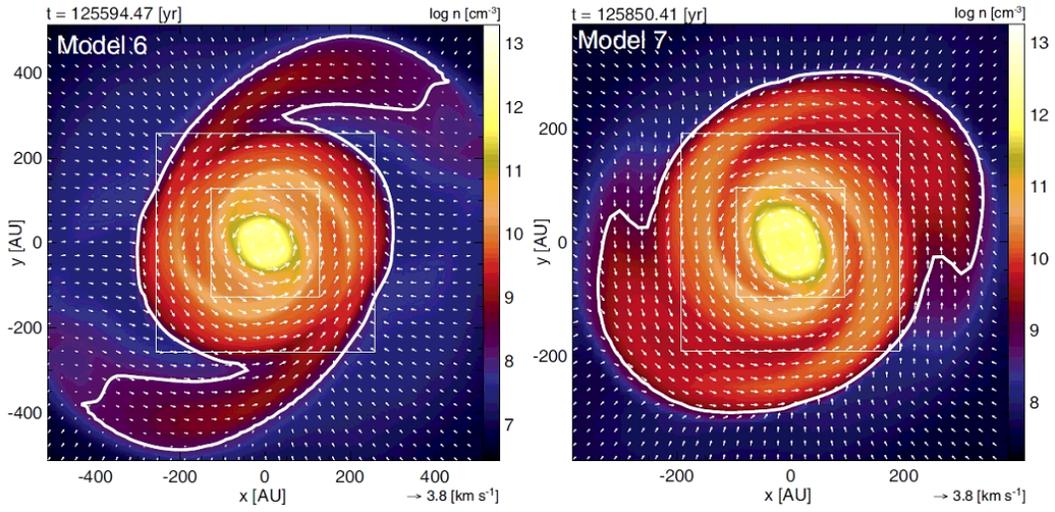}
\end{center}
\caption{
Density distribution ({\it color and contours}) and velocity vectors ({\it arrows}) on the equatorial plane for models 6 and 7.
The elapsed time $t$ is described in each panel.
}
\label{fig:12}
\end{figure}

Figure~\ref{fig:11} upper panel shows that at the same epoch after the circumstellar disk formation, the model with rapid initial rotation has a relatively large size disk.
In addition, the disk size begins to oscillate from the earlier evolution phase in the cloud with faster rotation.
For example, model 6 begins to show an oscillation at $\tc\sim2\times10^4$\,yr, while model 7 shows it for $\tc\gtrsim5\times10^4$\,yr.
The size oscillation occurs after the spiral structure develops in the circumstellar disk.
The spiral structure appears after the circumstellar disk grows sufficiently in size and mass \citep{machida10a}.
Figure~\ref{fig:12} shows the density distribution for models 6 and 7 on the equatorial plane after the circumstellar disk becomes more massive than the infalling envelope.
The figure indicates that the circumstellar disks show a clear spiral structure with a size of $300-400$\,AU.

Figure~\ref{fig:11} lower panel shows that the cloud with faster initial rotation creates a more massive circumstellar disk.
Thus, in such a cloud, the spiral structure and size oscillation appear in an earlier phase of the disk evolution.
In addition, the cloud with faster initial rotation has a less massive protostar before the size oscillation because mass accretion from the circumstellar disk onto the protostar is suppressed by a strong centrifugal force in such a model.
After the size oscillation occurs, the protostar unsteadily acquires its mass from the circumstellar disk, as described in \S\ref{sec:typical}.
Finally, the disk radius in each model increases rapidly for $\tc\gtrsim (0.7-0.9)\times 10^4$\,yr to reach $\sim300-400$\,AU (see, Table~\ref{table:2}).
This is because, by this epoch,  the circumstellar disk mass becomes greater than the protostellar mass, and magnetic braking becomes ineffective.

\subsubsection{Effect of Initial Cloud Masses}
\begin{figure}
\begin{center}
\includegraphics[width=140mm]{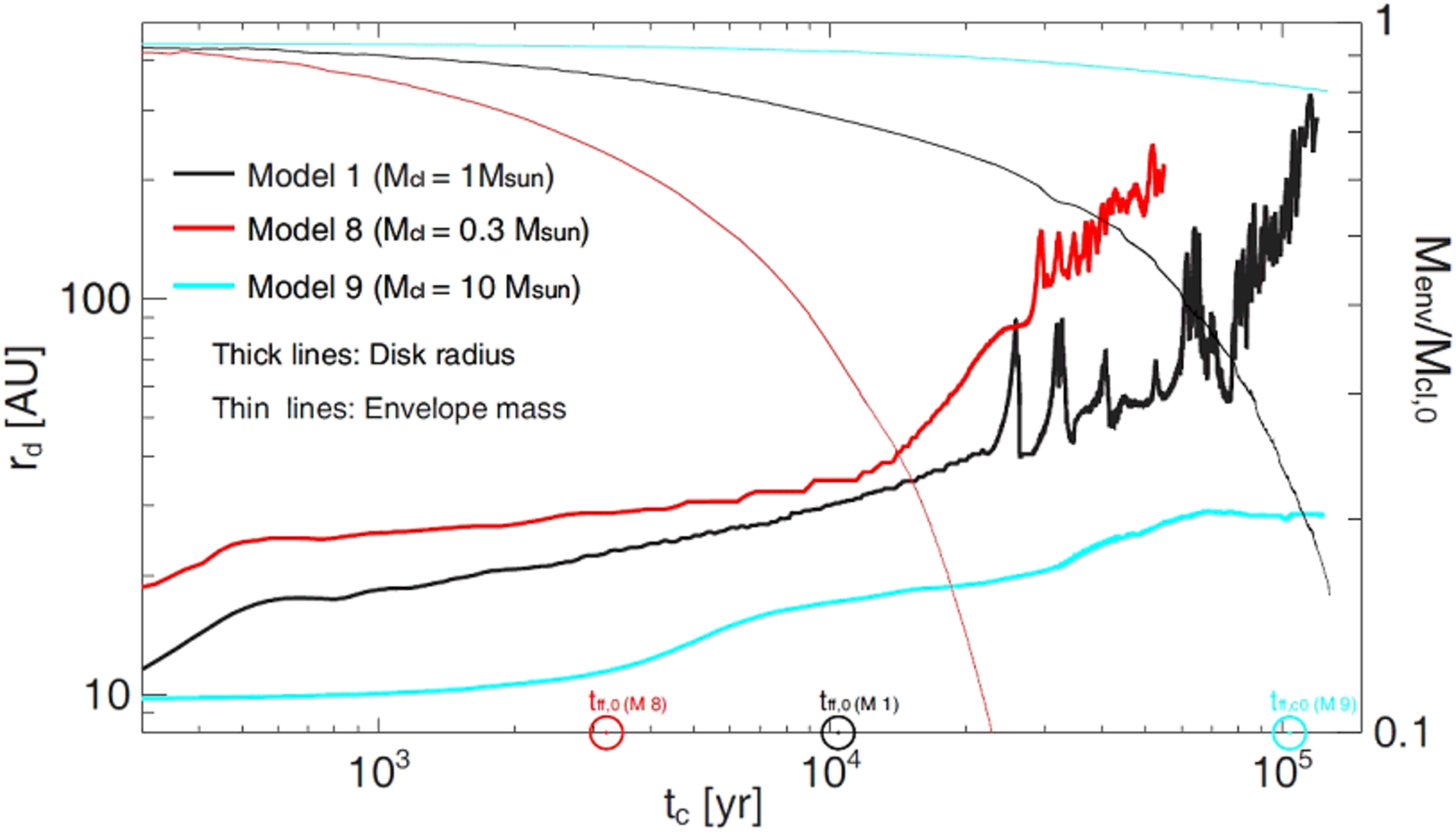}
\end{center}
\caption{
Disk radius (left axis) and mass of the infalling envelope normalized by the initial cloud mass are plotted against the elapsed time after the circumstellar disk formation for models 1, 8 and 9.
The freefall timescale for each model are noted by the circle on the horizontal axis.
}
\label{fig:13}
\end{figure}

Figure~\ref{fig:13} plots the disk radius and mass of the infalling envelope normalized by the initial cloud mass for models 1, 8, and 9 against the elapsed time after the circumstellar disk formation.
Each of these models has a different initial cloud mass but the same $\lambda$ and $\beta_0$.
The figure shows that the initially less massive cloud has a larger-size disk at the same epoch after the circumstellar disk formation.
The disk radii for models 1 and 8 rapidly increase after the mass of the infalling envelope deceases to $<20-40$\% of the initial cloud mass.
For model 8, which has an initial cloud mass of $M_{\rm cl}=0.3\msun$, the mass of the infalling envelope normalized by the initial cloud mass decreases to $\sim0.2$ at $\tc\simeq10^4$\,yr, and the circumstellar disk radius rapidly increases for  $\tc\gtrsim10^4$\,yr and reaches $\sim220$\,AU by the end of the calculation.
On the other hand, for model 9, which has an initial cloud mass of $M_{\rm cl}=10\msun$, the size of circumstellar disk does not exceed $30$\,AU by the end of the calculation ($\tc\simeq10^5$\,yr).
Thus, the disk size for model 9 is less than one-tenth of that for model 1 at $\tc\simeq10^5$\,yr.
Note that, for model 9,  because the freefall timescale at the cloud boundary is as large as $\sim10^6$\,yr, we cannot calculate the cloud evolution until the infalling envelope is depleted.

It is expected that the difference in the disk size for models with different initial cloud mass is caused by the remaining mass of the infalling envelope.
As described in \S\ref{sec:intro}, the massive infalling envelope brakes the circumstellar disk through the magnetic field lines.
The initial cloud mass and freefall timescale at the center of the cloud or at the cloud boundary are different in models 1, 8, and 9.
Thus, at the same epoch, a model with an initially massive cloud has a massive infalling envelope that can brake the circumstellar disk and suppresses the growth of the circumstellar disk by magnetic braking.
It is difficult to quantitatively relate the disk size and the mass of the infalling envelope among these models, because the initial distribution of the mass-to-flux ratio (see Fig.~\ref{fig:2}) and angular velocities (see Table~\ref{table:1}) are different in these models. 
However, Figure~\ref{fig:1} indicates that after the mass of the infalling envelope decreases sufficiently, the effect of magnetic braking weakens and a large-size circumstellar disk can form even in a strongly magnetized cloud.
Conversely, large-disk formation is  suppressed by magnetic braking when a massive infalling envelope remains around  the protostar.

\section{Discussion and Conclusion}
\label{sec:diss}
In this paper, to investigate the formation of the  circumstellar disk in a realistic setting, we set a dense molecular cloud core embedded in a low-density interstellar medium in a large simulation box with size 16 times the radius of the molecular cloud core, in which a uniform magnetic field parallel to the rotation axis is adopted.
In the gravitationally collapsing cloud, it is expected that the disk formation is suppressed by magnetic braking which effectively transfers the angular momentum from the circumstellar disk, as described in past studies \citep{mellon08,mellon09,hennebelle09,duffin09}.
In reality, our results indicate that the circumstellar disk formed in a magnetized cloud is much smaller than that in an unmagnetized cloud (see \S\ref{sec:mags}).
However, we also showed that the circumstellar disk exceeding $\sim100$\,AU can form even in a strongly magnetized cloud ($\lambda\sim1$), which contradicts previous studies.
The past studies predicted that magnetic braking suppresses the disk formation and limits the size of the circumstellar disk within $\sim10$\,AU.
This inconsistency is attributed to the initial setting of the host cloud and the calculation time after the circumstellar disk formation.
In this study, we calculated the cloud evolution until almost all gas inside the initial cloud falls onto either the circumstellar disk or the protostar.
In other words, we execute the calculation until the freefall timescale at the cloud boundary is exceeded.
As described in \S\ref{sec:intro}, as the cloud mass is depleted, magnetic braking is expected to be ineffective because the less massive infalling envelope cannot brake the circumstellar disk.
In \S\ref{sec:results}, we numerically showed that magnetic braking weakens with time and a large-size circumstellar disk can form. 
In this section, we analytically discuss the timescale of magnetic braking and disk formation.

The angular momentum of the cloud or disk is transferred into the external medium through a torsional Alfv$\acute{\rm e}$n wave.
Magnetic braking becomes effective when the angular momentum transferred into the external medium is comparable to that of the rotator (e.g., the molecular cloud or disk).
In other words, magnetic braking becomes effective when the moment of inertia of the gas through which the Alfv$\acute{\rm e}$n waves have propagated becomes comparable to the moment of inertia of the cloud or disk.
%% \citep{tajima02}.
With this condition, the magnetic braking timescale can be described as 
\begin{equation}
\tau_{||} \simeq \dfrac{z_d}{v_{\rm A, ex}} \dfrac{\rho_d}{\rho_{\rm ex}},
\label{eq:mb}
\end{equation}
where $z_d$, ${v_{\rm A, ex}}$, $\rho_d$ and $\rho_{\rm ex}$ are the half thickness of the cloud (or the disk), Alfv$\acute{\rm e}$n velocity in the external medium, cloud (or disk) density, and density of the external medium, respectively \citep{mestel56,nakano72,mouschovias80}.
Thus, magnetic braking becomes effective when a typical timescale of the system ($\tau$) is longer than the magnetic braking timescale $\tau > \tau_{||}$.
With equation~(\ref{eq:mb}), the condition for magnetic braking becoming effective is described as
\begin{equation}
\tau v_{\rm A,ex}\, \rho_{\rm ex} \gtrsim z_d\, \rho_d.
\label{eq:mb2}
\end{equation}
The left-hand side of equation~(\ref{eq:mb2}) corresponds to the line density of the external medium in the range at which the Alfv$\acute{\rm e}$n wave can reach in $\tau$, while the right-hand side corresponds to the line density of the cloud or disk.
Equation~(\ref{eq:mb2}) indicates that magnetic braking is effective only when the mass of the external medium is greater than the mass of the rotator.
In other words,  magnetic braking does not take place when sufficient mass does not exit around the rotator.
This is natural because a less massive external medium cannot brake a massive object.
This condition can be also understood from equation~(\ref{eq:mb}).
When an uniform magnetic field  is assumed, equation~(\ref{eq:mb}) indicates that the magnetic braking timescale is proportional to $\tau_{\vert \vert}\propto \rho_{\rm ext}^{-1/2}$.
Note that the Alfv$\acute{\rm e}$n speed is proportional to $\rho_{\rm ext}^{-1/2}$.
Thus, the magnetic braking timescale becomes longer as the external density decreases.
For example, when the external density is extremely low, magnetic braking does not take place.

We estimate the magnetic braking timescale for model 1 (initially the strongest magnetic field model).
First, we estimate the magnetic-braking timescale of the molecular cloud core embedded in the interstellar medium.
For model 1, the density contrast between the center of the cloud core and the external medium is 82 at the initial state, and  Alfv$\acute{\rm e}$n speed in the external mediums is $v_{\rm A,ext}=1.3$\,km\,s$^{-1}$.
Thus, with equation~(\ref{eq:mb}), the magnetic braking timescale becomes $\tau_{\rm ||} \simeq 8\times10^5$\,yr, which is much longer than the freefall timescale of the cloud ($t_{\rm ff,c}=1.04\times10^4$\,yr) or our calculation time ($\simeq10^5$\,yr).
Thus, it is expected that the cloud rarely loses its angular momentum by magnetic braking of the interstellar medium.
In \S\ref{sec:typical}, we showed that the total angular momentum of the cloud core is rarely decreased through the calculation.

Next, we estimate the magnetic-braking timescale of the circumstellar disk.
First, we assume that the circumstellar disk exists only in the low-density interstellar medium (thus, not including the infalling envelope) to simply estimate the effect of the low-density interstellar medium outside the gravitational sphere.
The density contrast between the circumstellar disk ($n\simeq10^{10}\cm$) and interstellar medium ($3\times10^4\cm$) is about $3 \times 10^6$, and the circumstellar disk has a scale height of $\sim 10$\,AU.
With these parameters, the magnetic braking timescale of the circumstellar disk is $\tau_{||}\simeq1.1\times10^8$\,yr.
Thus, the interstellar medium cannot brake the circumstellar disk in our setting.
Note that, in this study, we adopted a somewhat higher density of the interstellar medium ($3\times10^4\cm$) than the observed value of the interstellar medium to limit the  Alfv$\acute{\rm e}$n speed in the interstellar medium, because it is difficult to calculate the cloud evolution over a long time with a large Alfv$\acute{\rm e}$n speed.
Thus,  the magnetic braking timescale in the actual interstellar medium is longer than in our setting with a lower external density, because it is proportional to $\rho_{\rm ext}^{-1/2}$.
In summary, magnetic braking from the low-density interstellar medium is completely ignored in the formation and evolution of the circumstellar disk.

On the other hand, the infalling envelope in the region of $r<2 r_{\rm BE}$ can brake the circumstellar disk in the main accretion phase.
The torsional Alfv$\acute{\rm e}$n wave generated by the center of the cloud can reach the cloud boundary ($r=5400\,$AU) in $\sim2\times10^5$\,yr ($\sim2\,t_{\rm ff,b}$), where the Alfv$\acute{\rm e}$n speed of the external medium $v_{\rm A,ext}=1.3$\,km\,s$^{-1}$ is adopted.
Note that, in reality, because the magnetic field is amplified, the torsional Alfv$\acute{\rm e}$n wave can reach the cloud boundary in several freefall timescales.
Thus, when the infalling envelope is more massive than the circumstellar disk, the angular momentum of the circumstellar disk is effectively transferred by magnetic braking.
In the main accretion phase, however,  the mass of the infalling envelope decreases with time.
Thus, it is expected that magnetic braking gradually becomes ineffective with time.
However, it is difficult to estimate the magnetic braking timescale of the infalling envelope quantitatively, because the infalling envelope drastically changes its density distribution in a several free-fall timescale that corresponds to  the timescale for the circumstellar disk formation.

Thus, the disk formation process and final size of the circumstellar disk cannot be investigated using a simple analytical framework.
Only a less-massive and small-size circumstellar disk may appear in a magnetized cloud when magnetic braking significantly transfers the angular momentum by the end of the main accretion phase.
In this study, however, we calculated the cloud evolution by the end of the main accretion phase and showed that the circumstellar disk exceeding $\sim100$\,AU can form even in a strongly magnetized cloud.
In a magnetized cloud, magnetic braking becomes ineffective and the disk size increases rapidly in the later main accretion phase.
It is natural that magnetic braking only redistributes the angular momentum from the circumstellar disk into the infalling envelope.
Because the infalling envelope finally falls onto the circumstellar disk, the excess angular momentum gives rise to a rotation supporting the disk in the later main accretion phase.
Instead of magnetic braking, only the protostellar outflow can transfer the angular momentum to the interstellar medium, because the gas with a certain amount of angular momentum is directly ejected into the interstellar medium from the circumstellar disk by the protostellar outflow.
Because the protostellar outflow is powered by mass accretion onto the circumstellar disk, the outflow disappears in the later main accretion phase, which also corresponds to rapid growth of the circumstellar disk.
Thus, the circumstellar disk begins to grow after the protostellar outflow sufficiently weakens or disappears.
In addition, in the later main accretion phase, the spiral structure appearing in the circumstellar disk can contribute to the redistribution of the angular momentum in the circumstellar disk.
Thus, it is difficult to specify which mechanism is effective for the angular momentum transfer in a magnetized collapsing cloud.
However, the large-size disk comparable to observations can form in a strongly magnetized cloud.

In this study, we mainly investigated the evolution of the collapsing cloud with a mass of $1\msun$ in the initial state and found that the circumstellar disk with a size of several hundred AU appears in the later main accretion phase.
On the other hand, we cannot calculate the cloud evolution for an initially massive cloud until the infalling envelope disappears.
However, a massive cloud has a larger angular momentum (or larger centrifugal radius).
Thus, a circumstellar disk much larger than $\sim100$\,AU may form in an initially massive cloud.

Finally, we comment on the geometry of the initial magnetic field.
In this study, we assumed the magnetic field lines to be parallel to the rotation axis.
With ideal MHD calculation, \citet{hennebelle09} already reported the formation of the circumstellar disk when the magnetic field lines are inclined from the rotation axis.
Thus, we need to investigate the inclined field case with non-ideal MHD effects in the future.
%%Thus, a large-size disk appears even in the early main accretion phase with inclined field lines.

\acknowledgments
%%MNM have greatly benefited from discussions with ~T. Kudoh.
Numerical computations were carried out on NEC SX-9 at Center for Computational Astrophysics, CfCA, of National Astronomical Observatory of Japan, and NEC SX-8 at the Yukawa Institute Computer Facility.
This work was supported by the Grants-in-Aid from MEXT (20540238, 21740136).

\clearpage
%%%%%%%%%%%%%
%%% Table %%%
%%%%%%%%%%%%%
\begin{table}
\setlength{\tabcolsep}{3pt}
\caption{Model parameters}
\label{table:1}
\footnotesize
\begin{center}
%%\scalebox{.5}{%
\begin{tabular}{c|cccccccccccc} \hline
{\footnotesize Model} & $n_{\rm c,0}$ {\scriptsize [$\cm$]} & $r_{\rm cl}$ {\scriptsize [AU]} & $M_{\rm cl}$  {\scriptsize [$\msun$]}  &
 $\alpha$ & $\omega$ & $\lambda$ &$B_{0}$ {\scriptsize [$\mu$G]}& $\Omega_0$ {\scriptsize [s$^{-1}$]}  & $\beta_0$ & $R_{\rm cf}$ {\scriptsize [AU]} & R/I$^*$ 
\\ \hline
1     & $3\times 10^6$ & 5400  & 1   & 1.0 & 0.1 & 1  & 176 & $2.3\times10^{-13}$  & 0.03 & 2150   & R \\
2     & $3\times 10^6$ & 5400  & 1   & 0.3 & 0.1 & 2  & 92  & $2.3\times10^{-13}$  & 0.03 & 2150   & R\\
3     & $3\times 10^6$ & 5400  & 1   & 0.1 & 0.1 & 3  & 58  & $2.3\times10^{-13}$  & 0.03 & 2150   & R\\
4     & $3\times 10^6$ & 5400  & 1   & 1.0 & 0.1 & 1  & 176 & $2.3\times10^{-13}$  & 0.03 & 2150   & I \\
5     & $3\times 10^6$ & 5400  & 1   & 0   & 0.1 & $\infty$  & 0  & $2.3\times10^{-13}$  & 0.03  & 2150 & R\\
6     & $3\times 10^6$ & 5400  & 1   & 1.0  & 0.15 & 1  & 176 & $4.7\times10^{-13}$  & 0.04 & 4800 & R\\
7     & $3\times 10^6$ & 5400  & 1   & 1.0 & 0.05 & 1 & 176 & $1.2\times10^{-13}$  & 0.007 & 540   & R\\
8     & $3\times 10^7$ & 1740  & 0.3 & 1.0 & 0.1 & 1  & 566 & $7.4\times10^{-13}$  & 0.03 & 680    & R \\
9     & $3\times 10^4$ & 54000 & 10 & 1.0 & 0.1 & 1  & 17.6 & $2.3\times10^{-14}$  & 0.03 &21500    & R \\
\hline
\end{tabular}
\end{center}
$^*$ R: Resistive MHD model, I: Ideal MHD model
\end{table}

%%%%%%%%%%%%%
%%% Table %%%
%%%%%%%%%%%%%
\begin{table}
\setlength{\tabcolsep}{3pt}
\caption{Calculation results}
\label{table:2}
\footnotesize
\begin{center}
%%\scalebox{.5}{%
\begin{tabular}{c|cccccccccccc} \hline
{\footnotesize Model} &  $t$ {\scriptsize [$t_{\rm ff,c}$]} & $t_0$ {\scriptsize [$t_{\rm ff,c}$]} & $r_{\rm d}$  {\scriptsize [AU]}  &
 $M_{\rm disk}$ {\scriptsize [$\msun$]} 
%%& $M_{\rm ps}$ {\scriptsize [$\msun$]} & $M_{\rm env}$ {\scriptsize [$\msun$]} 
\\ \hline
1 & 12.8    & 3.39   & 290  & 0.49 \\
2 & 12.3    & 3.09   & 440  & 0.39  \\
3 & 13.2    & 2.95   & 370  & 0.42 \\
4 & 11.2    & 3.40   & 120  & 0.31  \\
5 & 11.9    & 2.87   & 1200 & 0.91 \\
6 & 11.3    & 3.57   & 390  & 0.50 \\
7 & 11.4    & 3.38   & 410  & 0.52 \\
8 & 14.2    & 3.59   & 220  & 0.12  \\
9 & 4.41    & 3.37   & 29   & 0.05  \\
\hline
\end{tabular}

\end{center}

\end{table}

\clearpage
%%%%%%%%%%
% Fig. 1 %
%%%%%%%%%%

\clearpage
%%%%%%%%%%
% Fig. 2 %
%%%%%%%%%%

\clearpage
%%%%%%%%%%
% Fig. 3 %
%%%%%%%%%%

\clearpage
%%%%%%%%%%
% Fig. 4 %
%%%%%%%%%%

\clearpage
%%%%%%%%%%
% Fig. 5 %
%%%%%%%%%%

\clearpage
%%%%%%%%%%
% Fig. 6 %
%%%%%%%%%%

\clearpage
%%%%%%%%%%
% Fig. 7 %
%%%%%%%%%%

\clearpage
%%%%%%%%%%
% Fig. 8 %
%%%%%%%%%%

\clearpage
%%%%%%%%%%
% Fig. 9 %
%%%%%%%%%%

\clearpage
%%%%%%%%%%
% Fig. 10%
%%%%%%%%%%

\clearpage
%%%%%%%%%%
% Fig. 11%
%%%%%%%%%%

\clearpage
%%%%%%%%%%
% Fig. 12%
%%%%%%%%%%

\clearpage
%%%%%%%%%%
% Fig. 13%
%%%%%%%%%%

\end{document}